# UAV-Based Remote Sensing of Soil Moisture Across Diverse Land Covers: Validation and Bayesian Uncertainty Characterization

Runze Zhang, Ishfaq Aziz, Derek Houtz, Yuxiang Zhao, Trent W. Ford, Adam C. Watts, and Mohamad Alipour

***Abstract*— High-resolution soil moisture (SM) observations are critical for agricultural monitoring, forestry management, and hazard prediction, yet current satellite passive microwave missions are unable to directly provide retrievals at tens-of-meter spatial scales. Unmanned aerial vehicle (UAV)–mounted microwave radiometry presents a promising alternative, but most evaluations to date have focused on agricultural settings, with limited exploration across other land covers and few efforts to quantify retrieval uncertainty. This study addresses both gaps by evaluating SM retrievals from a drone-based Portable L-band Radiometer (PoLRa) across shrubland, bare soil, and forest strips in Central Illinois, U.S., using a 10-day field campaign in 2024. Controlled UAV flights at altitudes of 10 m, 20 m, and 30 m were performed to generate brightness temperatures ($T_B$) at spatial resolutions of 7 m, 14 m, and 21 m. SM retrievals were carried out using multiple tau-omega-based algorithms, including the single channel algorithm (SCA), dual channel algorithm (DCA), and multi-temporal dual-channel algorithm (MT-DCA). A Bayesian inference framework was then applied to provide probabilistic uncertainty characterization for both SM and vegetation optical depth (VOD). Results show that the gridded $T_B$ distributions consistently capture dry-wet gradients associated with vegetation density variations, and spatial correlations between polarized observations are largely maintained across scales. Validation against *in situ* measurements indicates that PoLRa-derived SM retrievals from the SCA-V and MT-DCA algorithms achieve unbiased root-mean-square errors (ubRMSE) generally below 0.04 m³/m³ across different land covers. Bayesian posterior analyses confirm that reference SM values largely fall within the derived uncertainty intervals, with mean uncertainty ranges around ± 0.02 m³/m³ and ± 0.11 m³/m³ for SCA and DCA-related retrievals. These findings underscore the potential of UAV-mounted PoLRa for high-resolution SM retrieval across varied landscapes and emphasize the need for standardized calibration and uncertainty quantification frameworks to support broader scientific and operational adoption.***

This work was supported by the United States Department of Agricultural Forest Service under Grant 110789. *(Corresponding author: Runze Zhang).*

Runze Zhang, Ishfaq Aziz, Yuxiang Zhao, and Mohamad Alipour are with the Department of Civil and Environmental Engineering, University of Illinois Urbana-Champaign, Urbana, IL 61801 USA (e-mail: runze@illinois.edu; ishfaqa2@illinois.edu; zhao132@illinois.edu; alipour@illinois.edu).

Derek Houtz is with Microwave Remote Sensing Group of Swiss Federal Institute for Forest, Snow, and Landscape Research, Zürich 8903 Switzerland (e-mail: derek.houtz@wsl.ch).

Adam C. Watts is with the Pacific Wildland Fire Sciences Laboratory, United States Forest Service, Wenatchee, WA 98801 USA (e-mail: adam.watts@usda.gov).

Trent W. Ford is with Illinois State Water Survey, Prairie Research Institute, University of Illinois, Urbana-Champaign, Champaign, IL 61820 USA (e-mail: twford@illinois.edu).

supplemental materials

## I. INTRODUCTION

SOIL moisture (SM) plays a foundational role in hydrology and land surface water cycle, serving as either an initial condition or a boundary condition in a wide range of hydrological models [1]. Accurate monitoring of SM dynamics is essential for advancing our fundamental understanding of land-atmosphere interactions and hydrological processes, thereby supporting progress across numerous Earth and environmental science disciplined [2-4]. In addition, previous studies have demonstrated that timely and precise SM information can significantly improve a variety of applications, including weather forecasting, drought monitoring, agricultural productivity assessments, food security prediction, and climate modeling [5-8]. Given the importance of SM for both fundamental scientific research and practical applications, there is a critical need for observations that offer SM of both high precision and scalable spatial resolutions.

Microwave radiometry at lower frequencies, particularly at L-band (~ 1.4 GHz), offers significant advantages for SM monitoring due to the strong dielectric contrast between water and soil. Compared to C- and X-bands, L-band measurements exhibit superior atmospheric penetration, reduced vegetation attenuation, and greater sensing depth (~ 5 cm) [9]. The accuracy and applicability of satellite-based L-band SM products have been extensively validated across a variety of studies [10-14]. However, their coarse spatial resolutions, typically on the order of tens of kilometers, are primarily suited for large-scale applications such as numerical weather prediction and climate modeling [15]. Currently, while SM observations span scales from *in situ* point measurements to satellite footprints (~ 40 km), there remains a critical gap at intermediate scales (particularly from tens to hundreds of meters), significantly constraining the ability to capture localized variability and spatial hierarchies of SM.

To address these resolution gaps, aircraft-based L-band radiometers have been used to provide SM with both high accuracy and fine spatial resolutions [16-18]. However, airborne campaigns typically require complex logistics and entail substantial costs. As an alternative, various spatial downscaling schemes, such as the trapezoid model, have been extensively studied and applied to passive microwave SM products, aiming to preserve the accuracy of radiometer-based retrievals while capturing fine-scale spatial variability [19-21]. Nevertheless, these downscaled products often face significant validation challenges due to the scarcity of high-resolution reference data that can simultaneously provide accurate magnitude of soil water content and correctly represent the spatial patterns of SM variations nested within the coarse-scale pixel units. Additionally, the risk of introducing errors from ancillary datasets limits the reliability of downscaled SM fields, suggesting that directly capturing high-resolution SM from microwave observations remains the optimal strategy at present. Although spaceborne active microwave sensors, such as synthetic aperture radar (SAR), can deliver higher spatial resolution, their limited revisit frequency (~ 12 days), heighted sensitivity to surface roughness changes, higher operating frequency, and lower effective measurement depth, partially restrict their effectiveness for applications requiring timely and accurate SM retrievals [1, 22].

In recent years, the combination of unmanned aerial vehicles (UAVs) and microwave radiometers has emerged as a promising approach to capture SM at scalable resolutions, particularly ranging from a few meters to several hundred meters [23-25]. UAV-mounted L-band radiometers offer not only adjustable high spatial resolutions but also retrieval accuracy, cost-effectiveness, and flexible observation scheduling. The Portable L-band Radiometer (PoLRa), developed by TerraRad Tech, represents a novel, compact L-band radiometer specifically designed for the deployment on various mobile platforms [26]. Unlike other small-scale L-band radiometers, such as the lobe differencing correlation radiometer (LDCR) and the airborne radiometer at L-band (ARIEL), which are limited to single-polarization measurements, PoLRa is capable of acquiring off-nadir brightness temperatures ($T_B$) at both vertical and horizontal polarizations [26-28]. SM retrievals based on PoLRa installed on pole-, rail-, and vehicle-fixed platforms have demonstrated its ability to capture SM variations over bare soils and grasslands at meter-scale resolutions with reasonable accuracy [29-31]. However, compared to ground-based deployments, UAV-based PoLRa SM retrievals introduce additional challenges, such as quickly changing incidence angles and footprint scaling with height above ground, which have not yet been systematically evaluated.

Hence, the primary objective of this study is to provide a comprehensive evaluation of UAV-based PoLRa-derived SM estimates. The urgent need to quantify topsoil water content for agricultural management has led most validation studies of drone-based passive microwave SM retrievals to focus on croplands [23, 25]. Many hydrometeorological applications, such as forest harvest and drought prediction, require SM information across diverse vegetation conditions. Therefore, in this study, the performance of PoLRa-derived SM retrievals was assessed across natural shrublands, bare soils, and mixed forests during a 10-day field campaign conducted in Central Illinois. Following experiment protocols proposed by [18], the spatial consistency of SM across different resolutions was examined at three separate altitudes: 10 m, 20 m, and 30 m. Additionally, this study uniquely integrates surface temperature measurements acquired from the thermal infrared sensors onboard the Da-Jiang Innovations (DJI) Mavic 3T [32]. By separately considering canopy temperatures from thermal infrared observations and soil temperatures from *in situ* probes at approximately 5 cm depth, the retrieval framework becomes independent of the thermal equilibrium assumption typically valid for early morning observations [9]. This approach lays the groundwork for extending drone-based SM sensing beyond the conventional 6 a.m. observation window, enabling flexible diurnal measurements and facilitating the coordinated use of multiple UAV-mounted sensors for SM sensing.

Given the increasing need for real-time vegetation monitoring and the potential to infer fuel moisture from microwave-derived vegetation optical depth (VOD), the utilization of dual-polarized observations for simultaneously retrieving SM and VOD have become an area of growing interest [33, 34]. However, conventional dual-channel algorithm (DCA) is susceptible to large retrieval biases because of minor observational noises, imperfect forward models, compensatory effects during optimization, and the strong dependence between polarized observations [35, 36]. To mitigate these issues, the multi-temporal dual-channel algorithm (MT-DCA) was developed, enabling an overdetermined retrieval system where the number of observations exceed the number of unknown parameters [37, 38]. A simplified version of the MT-DCA approach was incorporated in this study to preliminary assess utility potential to UAV-mounted microwave radiometer data.

Beyond providing a comprehensive evaluation of PoLRa-derived SM retrieval across different land covers, resolutions, algorithms, and thermal assumptions, a secondary objective of this work is to estimate the retrieval uncertainty through the coupled application of Bayesian inference and Markov Chain Monte Carlo (MCMC) simulations [39, 40]. The application of probabilistic distributions as uncertainty information for SM and VOD retrievals have not been widely explored, especially when compared to extensive efforts focused on their deterministic error characterizations. Previous uncertainty characterization efforts often involve assigning fixed ranges to input parameters and observations, these methods treat all values within the range as equally probable, making the results highly sensitive to initial guesses [17, 35]. Further, non-probabilistic inversions may not represent complex multimodal solutions or inversion ambiguities resulting from noisy real-world data. In contrast, Bayesian inference yields full posterior distributions that integrate both model-data mismatch and inversion ambiguities, providing a more robust representation of retrieval uncertainty. This study primarily focuses on demonstrating the feasibility of applying Bayesian inference to SM retrievals in the context of a zeroth-order radiative transfer model. The inclusion of credible intervals around the retrieval estimates further enhances the interpretability and reliability of the resulting SM and VOD products.

## II. Experiment Description

All field experiments were conducted at the Phillips Tract Research Plots within the University of Illinois Natural Areas, Urbana, Illinois (**Fig. 1a**). The study site spans approximately 240 m × 180 m and features seven parallel strips with varying land cover types, including shrublands, bare soil, and mixed forest (**Fig. 1b**). Shrub strips are managed under different vegetation clearing frequencies, ranging from annual to multiple-year intervals, resulting in distinct land cover characteristics despite comparable vegetation densities. For clarity, shrub strips are identified by a combination of the term 'Shrub' and their respective strip order (i.e., Shrub 1 for the bottom-most strip and Shrub 5 adjacent to bare soil) (**Fig. 1b**). *In situ* measurements of SM and soil temperature at a 5 cm depth were collected at 35 fixed locations, marked with blue triangles on the map, using METER TERO12 probes [41]. These 35 points were evenly distributed across forest, soil, and transitional shrubland areas. To ensure the measurement consistency, each location was permanently marked with red flags, and three replicate measurements were taken around each flag and averaged to represent local conditions for validation. However, due to accessibility constraints, several measurements were collected along pathways between shrub strips, potentially limiting their representativeness of adjacent strips. The experimental campaign was conducted from September 4 to 13, 2024, with flights and measurements restricted to a 9:30 – 11:30 a.m. window each day to minimize diurnal variability induced by temperature and evapotranspiration. Experiments on September 13 were curtailed due to unexpected rainfall. Despite this, soil samples had been collected on the final day (prior to precipitation) for gravimetric water content measurements and soil texture analysis, revealing an average clay fraction of 8.5% across the site.

To minimize battery replacements and optimize flight efficiency, PoLRa flights were conducted in two batches: 20 m and 30 m (first batch), and 10 m (second batch). The 30 m flight altitude was selected based on maximum tree heights (of around 26 m) in the forest strip. Therefore, observations at 10 m and 20 m do not include the contribution from forested areas. PoLRa scanned each strip from right to left and back (i.e., left to right) before moving to the next, with flight paths designed to maximize coverage of homogeneous areas while avoiding vegetation contact. Given PoLRa 3 dB full beamwidth of 37º, footprint dimensions along and across track were calculated, accounting for beam position correction and incidence angle variations. Although PoLRa is designed for observation at a nominal 40º, actual observation angles could span from 20º to 60º. Only data with incidence angles between 30º and 50º were retained for further analysis. To standardize spatial resolutions, conservative circular footprints were approximated at 7 m, 14 m, and 21 m in diameter for flight altitudes of 10 m, 20 m, and 30 m, respectively (**Fig. 2b**). The footprint sizes can be approximated as constant over the relatively fixed flight height-above-ground. Given the width of each strip, the use of these along-track sizes (i.e., width distance in **Fig. 2b**) as spatial resolutions might lead to misunderstanding that an individual emission observed over a strip is partly contributed by its adjacent strip, particularly with a flight elevation of 30 m. On the other hand, homogeneity and frequent sampling along the strip make the neighboring footprint-scale observations rather consistent, barely affecting the gridded $T_B$ aggregates and their subsequent SM retrievals. Cold-sky calibrations of PoLRa were conducted daily after each flight by directing the antenna towards the sky, with calibration procedures detailed in Section III. Additionally, a DJI Mavic 3T equipped with visible and infrared sensors was flown daily during the campaign, providing surface reflectance and canopy temperature data. The canopy temperatures estimated from

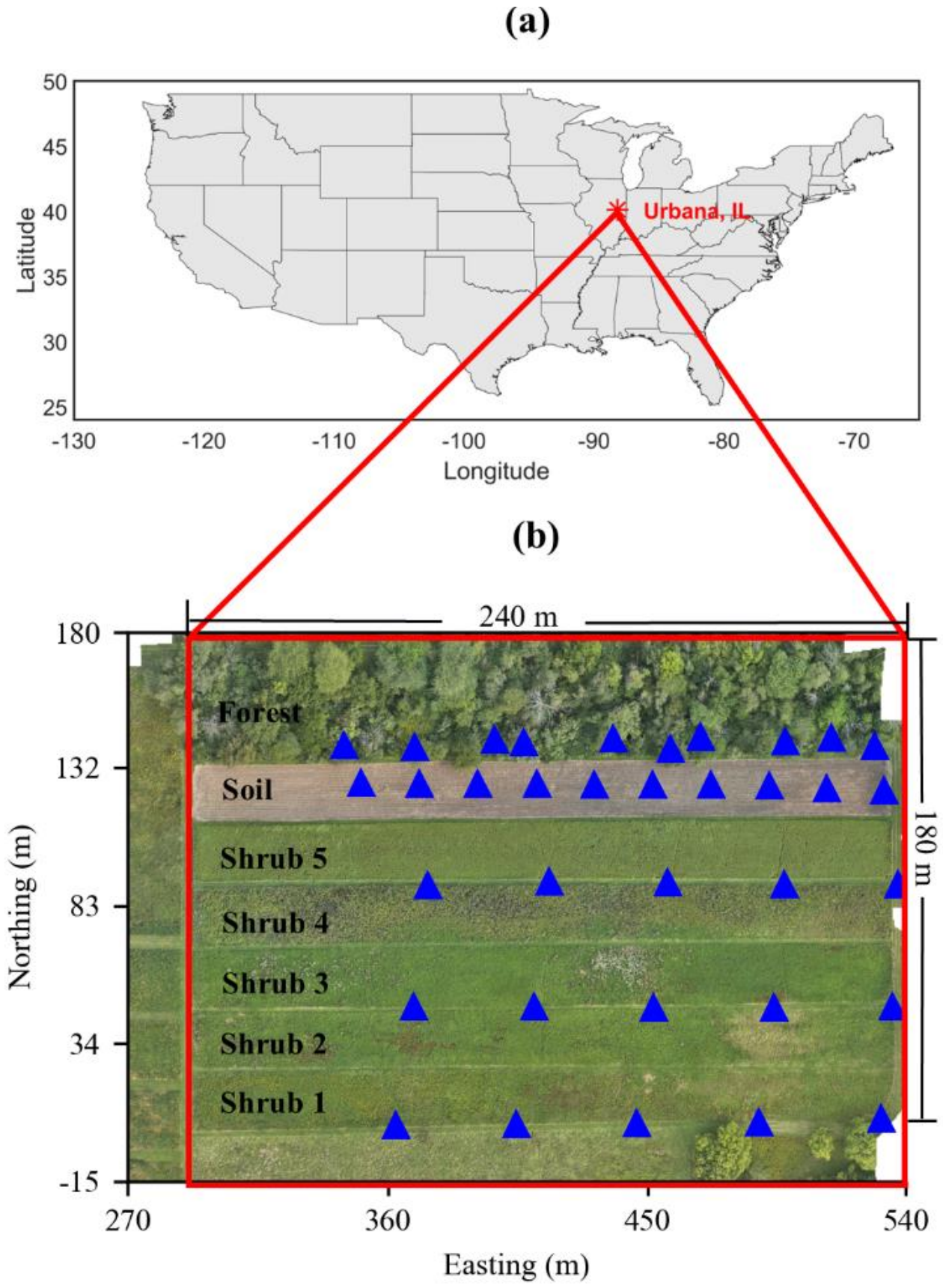

**Fig. 1.** (a) Geographic location of Urbana, Illinois on the U.S. map; (b) layout of the experimental site highlighted by the red box for evaluating soil moisture retrievals from the drone-mounted Portable L-band Radiometer (PoLRa), consisting of seven parallel strips covering forest, bare soil, and natural shrubland. Strip names are marked using bold font. Blue triangles denote the locations of *in situ* probe measurements.

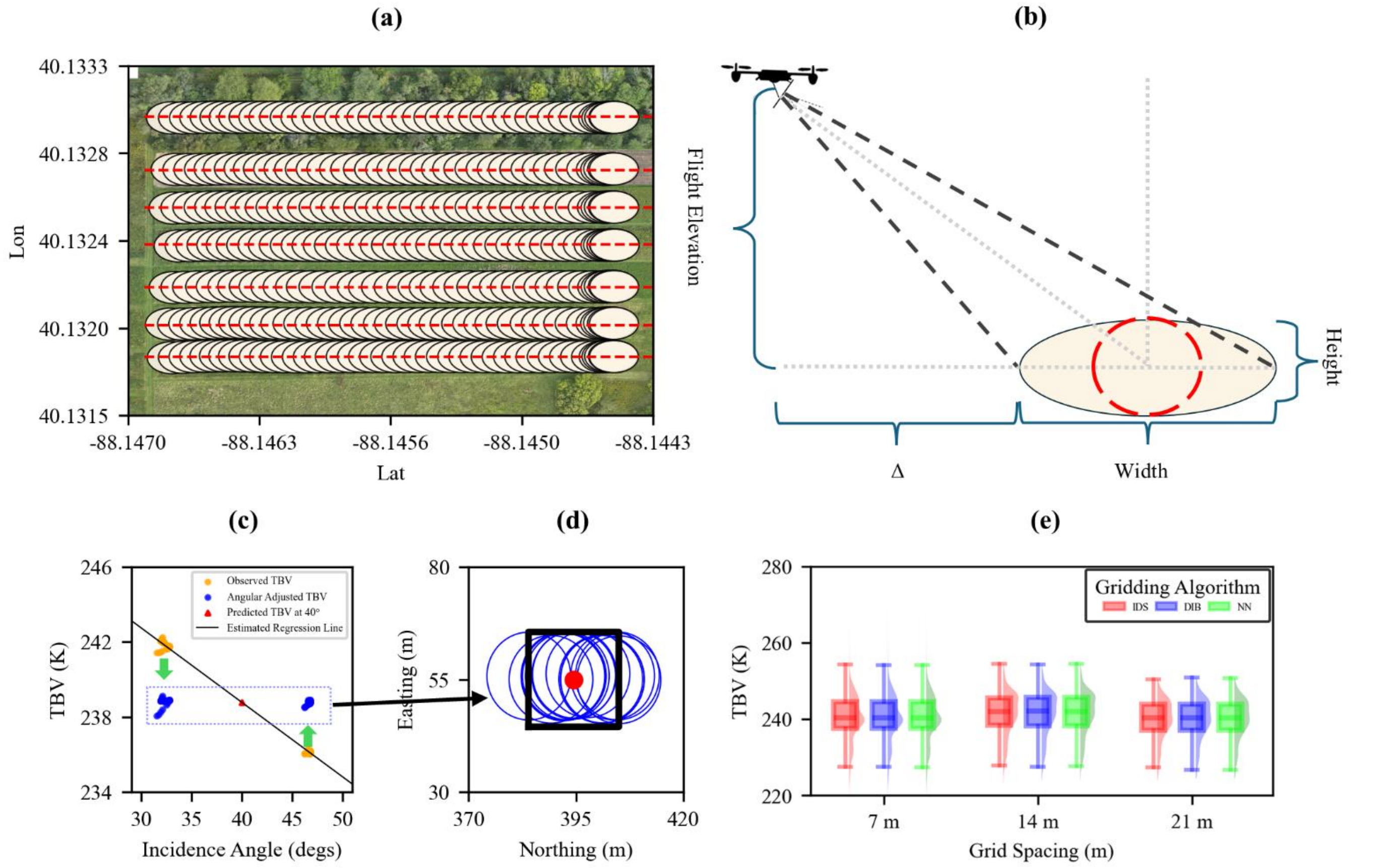

**Fig. 2**. (a) Spatial distribution of PoLRa 3 dB footprints along flight tracks at an altitude of 30 m; (b) conceptual geometry of one PoLRa 3 dB footprint-scale observation at an incidence angle of around 40º, where the red circle delineates the strategy for determining spatial resolution; (c) an example of angular normalization (to 40º) of all the footprint-scale observations of vertically-polarized brightness temperatures falling within the same pixel; (d) schematic illustrating the footprint-scale vertically-polarized brightness temperatures contained within a single 21 m pixel; (e) boxplots of vertically-polarized brightness temperatures at gridded scales of 7 m, 14 m, and 21 m, using three different spatial interpolation methods: inverse-distance-squared (IDS), drop-in-bucket (DIB), and nearest neighboring (NN).

Mavic 3T observations over vegetated strips were incorporated into the SM retrieval process to reduce dependence on the thermal equilibrium assumption, which is typically valid only near 6 a.m.

## III. Methodology

### *A. Sky Measurement Radiometer Characterization*

At microwave frequencies, the intensity of the observed emission is commonly represented by brightness temperature ($T_B$). In the context of PoLRa, quantifying the surface-emitted energy captured by the antenna ($T_{B_p}^{ant}$) begins with converting the raw voltage readings at the power detector ($u_p$) into the brightness temperature at the switch input ( $T_{B_p}^{in}$ ), which requires knowledge of radiometer gain (G) and inherent offset noise temperature ($T_{B_{off}}$) (1). Here, the subscript p refers to either horizontal (H) or vertical polarization (V). Both G and $T_{B_{off}}$ are not fixed over time and dynamic determined based on the relative changes between two internal reference measurements: the active cold source (ACS) and the ambient matched resistive source (RS) within the PoLRa. Using these internal references, G and $T_{B_{off}}$ can be calculated synchronously with $T_{B_p}^{in}$, as in (2) and (3). Given that the magnitudes of $T_{B_{RS}}$ and $T_{B_{ACS}}$ depend on the physical temperature of the internal reference sources and that the PoLRa hardware is not temperature-stabilized, it is essential to characterize the temperature dependence of $T_{B_{RS}}$ and $T_{B_{ACS}}$. For RS reference, $T_{B_{RS}}$ is usually assumed equivalent to its physical temperature ( $T_{RS}$ ), measured directly via internal sensor. In contrast, $T_{B_{ACS}}$ is expected to vary linearly with its physical temperature ($T_{ACS}$) (4) [42, 43].

$$T_{B_p}^{in} = G \cdot u_p + T_{B_{off}} \tag{1}$$

$$G = \frac{T_{B_{RS}} - T_{B_{ACS}}}{u_{RS} - u_{ACS}} \tag{2}$$

$$T_{B_{off}} = -G \cdot u_{RS} + T_{B_{RS}} \tag{3}$$

$$T_{B_{ACS}} = m \cdot T_{ACS} + b \tag{4}$$

It is important to note that the yielded $T_{B_p}^{in}$ comprises both 1) the portion of external radiation entering the antenna that reaches the switch input ($(1-\alpha_p)\cdot T_{B_p}^{ant}$) and 2) internal emissions from the transmission path (TP) composed for the antenna and cable ($\alpha_p \cdot T_{TP}$) (5). $T_{B_p}^{ant}$ represents the brightness temperature incident on the antenna at polarization $p$, primarily contributed by the target surface, which will serve as the observation inputs for SM and VOD retrievals. The parameter $\alpha_p$ is the absorption coefficient of the antenna-cable system and is derived from $L_p$ which denotes the cumulated loss between the antenna and switch input in decibels (dB) (6), resulting from non-ideal antenna efficiency, cable losses, and connector losses [26, 29]. In this study, polarization effects on antenna-cable losses are assumed negligible ($L_H = L_V = L$), and $T_{TP}$ is approximated as the average of the antenna ($T_{ant}$) and cable temperatures ($T_{cab}$) (7).

With voltage and temperature measurements, $T_{B_p}^{ant}$ can be derived once the parameters $m$, $b$, and $L$ are known (8). Calibration of these ACS-related parameters (m, b, and L) is typically achieved by introducing an external emission with known $T_B$ values and observing this known source over a range of physical temperatures. Sky and water surfaces are often used as reliable references for calibrating low-frequency radiometers [18]. At L-band, the sky $T_B$, including extraterrestrial radiation, is often on the order of several Kelvin [44, 45]. Assuming a sky $T_B$ of 5 K and using voltage measurements from PoLRa pointed at the clear sky, m, b, and L values can be determined via an iterative optimization procedure that minimizes the differences between $T_{B_p}^{ant}$ (sky) and the expected 5 K value. In this study, $m$, $b$, and $L$ parameters were adopted from prior calibration performed on the same PoLRa instrument (Table 1) [30]. Daily sky-pointing measurements during the field campaigns were additionally used to compute daily polarized temperature offsets ($d_p$), partially compensating for systematic deviations in $T_{B_p}^{ant}$ likely caused by impedance mismatches among antenna and cables or changes in background radiation (low-magnitude offsetable interference) (Table 1). Notably, the correction illustrated by (8) is only designed for this work, unlikely to be generalized to all PoLRa utilization.

$$T_{B_P}^{in} = (1-\alpha_p)\cdot T_{B_p}^{ant} + \alpha_p \cdot T_{TP} \quad (5)$$

$$\alpha_p = 1 - 10^{\frac{L_p}{10}} \quad (6)$$

$$T_{TP} = \frac{T_{ant} + T_{cab}}{2} \quad (7)$$

$$T_{B_P}^{ant} = \frac{T_{B_P}^{in} - \alpha_p \cdot T_{TP}}{(1-\alpha_p)} + d_p \quad (8)$$

Table 1. PoLRa calibration parameters (m, b, and L) used for the ACS characterization and daily polarized temperature offsets ($d_p$)

| Applicable Period | Parameters | | |
|---|---|---|---|
| | m | b (K) | L (dB) |
| September 4 – 12, 2024 | 0.355 | -90.000 | -0.100 |

| Applicable Date | $d_H$ (K) | $d_V$ (K) |
|---|---|---|
| September 4, 2024 | -52.478 | -55.992 |
| September 5, 2024 | -63.117 | -59.058 |
| September 6, 2024 | -66.694 | -57.914 |
| September 7, 2024 | -63.236 | -51.538 |
| September 8, 2024 | -65.903 | -48.817 |
| September 9, 2024 | -83.946 | -63.600 |
| September 10, 2024 | -62.923 | -51.155 |
| September 11, 2024 | -62.688 | -50.287 |
| September 12, 2024 | -68.990 | -59.229 |

### *B. Spatial Gridding of Footprint-Scale Brightness Temperatures*

Along the prescribed flight paths, the PoLRa's 3 dB elliptical footprints at a flight altitude of 30 m have been calculated and illustrated in **Fig. 2a**. With an integrated sampling cycle of 69 ms, meaning that the time required to sequentially record one round of voltage and temperature measurements at the ACS, RS, H-pol, and V-pol switches, PoLRa generates approximately 14 measurement sets per second, corresponding to 28 $T_{B_p}^{ant}$ values. Given this high observation density, footprint-level retrievals not only demand substantial computational resources but are also more susceptible to random noises. Furthermore, substantial overlap among adjacent footprints results in oversampling (of redundant information), limiting the contribution of unique observational content. Given these, footprint-scale observations are commonly aggregated onto spatial grids, producing effective brightness temperature per pixel ($T_{B_p}^{grid}$) [46]. Following this approach, $T_{B_p}^{ant}$ samples collected at flight altitudes of 10 m, 20 m, and 30 m were grouped, filtered, and gridded at corresponding pixel resolutions of 7 m, 14 m, and 21 m. Before gridding, the latitude and longitude of each beam center were corrected based on attitude angles and reprojected into the Universal Transverse Mercator (UTM) coordinate system. Observations with incidence angle outside the 30° to 50° were excluded from further analysis.

Three interpolation methods were applied to assign $T_{B_p}^{grid}$ values, which are (1) Drop-in-Bucket method (DIB), (2) Nearest Neighbor method (NN), and (3) Inverse-Distance-Squared method (IDS) [47]. For DIB, the arithmetic mean of $T_{B_p}^{ant}$ samples within a grid cell is assigned as the representative $T_{B_p}^{grid}$ whereas NN directly selects the footprint $T_{B_p}^{ant}$ closest to the center of the grid cell as $T_{B_p}^{grid}$. Regarding the IDS interpolation, the effective $T_{B_p}^{grid}$ value is calculated as a weighted average of all the footprint-scale $T_{B_p}^{ant}$ data within the grid cell, where the weights are inversely proportional to the square of the distance between each boresight location and the grid cell center (9). **Fig. 2e** shows that the resulting distributions of $T_{B_p}^{grid}$ values are similar across these methods. Given that the IDS interpolation offers a favorable balance between resolution degradation and noise reduction, it was selected for subsequent SM analyses [46]. However, it is important to note that interpolation does not fully eliminate biases in $T_{B_p}^{grid}$ values. In addition to random instrument noise, residual radiation contribution from regions outside the nominal 3 dB footprints remains a potential source of error.

$$T_{B_p}^{grid} = \frac{\sum_{i=1}^{N} \beta_i T_{B_p}^{ant_i}}{\sum_{i=1}^{N} \beta_i}$$
$$\beta_i = \frac{1}{d_i^2} \tag{9}$$
$$d_i = \sqrt{(x_i - x^{grid})^2 + (y_i - y^{grid})^2}$$

where the subscript i denotes the ordinal number of one snapshot within the predefined grid cell. $x_i$ and $y_i$ represent the coordinates of the beam center of the ith snapshot.

*C. The τ – ω Semi-Empirical Model*

The $\tau - \omega$ model, a zeroth-order radiative transfer model, simulates microwave thermal emissions from the land surface in form of $T_B$ by accounting for vegetation attenuation effects [48]. Neglecting the minor contribution of reflected downwelling atmospheric and extraterrestrial radiation, the upwelling emission from the land surface can be represented by three components, which are (1) upward soil emission attenuated by the vegetation layer, (2) direct vegetation emission, and (3) vegetation emission reflected off the soil surface (10) [9]. Conventional SM retrieval is performed by minimizing the differences between observed and simulated $T_B$ values through iteratively updating the input variables, i.e., SM and VOD (or $\tau$). The forward simulation begins by converting an estimated volumetric water content into dielectric permittivity using a dielectric mixing model. The Mironov model is applied in this study [49]. Polarized surface reflectivity ($r_{s_p}$) is then calculated from the dielectric constant via the Fresnel equations. To account for surface roughness effects, the H-Q-N semi-empirical roughness model is used to derive the rough surface reflectivity ($r_{r_p}$) (12) [50]. The radiative contribution of the canopy layer is incorporated, with attenuation strength determined by $\tau$, a composite parameter influenced by vegetation structures, water content, biomass, and canopy temperature [51, 52].

The specific objective functions for retrieval depend on the chosen algorithm and the targeted variables. In the classical single channel algorithm (SCA), SM is retrieved from a single polarization $T_B$, while $\tau$ is estimated independently using multi-spectral visible light vegetation indices, such as the Normalized Difference Vegetation Index (NDVI), as in (16) and (17) [9]. In DCA, both H-pol and V-pol $T_B$ measurements are used simultaneously, allowing the retrieval of both SM and $\tau$ by minimizing the sum of squared residuals between observation and simulated $T_{B_p}$ values (14). However, DCA retrievals are often highly sensitive to observational noises and model imperfections [53]. This sensitivity is primarily due to the mutual dependence between dual-polarized $T_B$ values within the same snapshot, which limits the degrees of independent information available for inversion [36]. To mitigate this issue, the MT-DCA was developed, using four $T_B$ observations from two successive scans to retrieve three unknowns where SM keeps varying and $\tau$ is assumed to remain constant for scans from the two consecutive days (15) [37, 38]. Given that MT-DCA has rarely been evaluated for UAV-mounted L-band radiometers that capture high-resolution variations and may suffer from calibration challenges, this study adopts MT-DCA alongside traditional SCA and DCA approaches based on the same set of observations. However, relative to the original MT-DCA, the implementation in this study has been simplified by prescribing a fixed value for the single scattering albedo ($\omega$). Specifically, $\omega$ is set as 0.08 for vegetated surfaces, corresponding to the median value of global $\omega$ map derived from SMAP MT-DCA outputs [37]. Additionally, following [37], the roughness parameter (h) of 0.13 was applied across all land cover types. For consistency, these h and $\omega$ values have been uniformly utilized for all retrieval algorithms.

$$\begin{aligned} T_{B_p}^{grid} &= \gamma\left(1 - r_{r_p}\right)T_s + (1-\gamma)(1-\omega)T_c \\ &\quad + \gamma r_{r_p}(1-\gamma)(1-\omega)T_c \end{aligned} \tag{10}$$

$$\gamma = e^{-\frac{\tau}{\cos\theta}} \tag{11}$$

$$r_{r_p} = [Q\, r_{s_p} + (1-Q) r_{s_q}] e^{-h\cos^2\theta} \tag{12}$$

$$J_{SCA}(sm_{ret}) = \left[T_{B_p}^{sim}(sm_{ret}) - T_{B_p}^{grid}\right]^2 \tag{13}$$

$$\begin{aligned} J_{DCA}(sm_{ret}, \tau_{ret}) &= \left[T_{B_V}^{sim}(sm_{ret}) - T_{B_V}^{grid}\right]^2 \\ &\quad + \left[T_{B_H}^{sim}(sm_{ret}) - T_{B_H}^{grid}\right]^2 \end{aligned} \tag{14}$$

$$\begin{aligned} &J_{MT-DCA}\left(sm_{ret}^{t_1}, sm_{ret}^{t_2}, \tau_{ret}\right) \\ &= \sum_{i=1}^{2} \left[T_{B_V}^{sim}\left(sm_{ret}^{t_i}\right) - T_{B_V}^{grid,t_i}\right]^2 \\ &\quad + \left[T_{B_H}^{sim}\left(sm_{ret}^{t_i}\right) - T_{B_H}^{grid,t_i}\right]^2 \end{aligned} \tag{15}$$

*D. Ancillary Datasets*

In SM retrieval using the $\tau - \omega$ model, the canopy temperature ($T_c$) is often assumed to approximate the soil effective temperature ($T_s$) under the assumption of thermal equilibrium, primarily due to the difficulties in simultaneously measuring $T_c$ during radiometric observations. However, given that the experiments in this study were conducted between 9:30 a.m. and 11:30 a.m., a period which canopy and soil temperatures typically diverge, the use of a single soil effective temperature to represent both near-surface soil and canopy temperatures is likely inappropriate [9, 54]. Therefore, a thermal infrared sensor onboard the DJI Mavic 3T, operating at wavelengths of 8 to 14 µm, was deployed to independently measure $T_c$ , immediately prior to PoLRa observations, assuming minimal temperature fluctuation during the experimental window. At a flight altitude of 30 m, the unit pixel size of the Mavic 3T thermal images is approximately 0.1 m (**Fig. 3d**). These thermal maps were reprojected into predefined grid cells of 7 m, 14 m, and 21 m via the DIB approach (**Fig. 3a - c**). Additionally, daily averages of soil temperatures measured at a depth of 5 cm using TEROS 12 probes were used as $T_s$ for SM retrieval. During the experimental period, the standard deviations of daily 5 cm soil temperatures were generally within 2 K. Differences between $T_c$ and $T_s$ are relatively stable over time, typically ranging from 0 to -4 K (**Fig. 3e**). In contrast, surface soil temperatures (measured by Mavic 3T) are consistently higher than 5 cm soil temperatures, with differences reaching approximately 8 K (**Fig. 3e**). These large deviations could be partially caused by the flight height of 30 m that exceeds the manufacturer's recommended maximum distance of 25 m between the thermal camera and the ground surface (private communication with the DJI support).

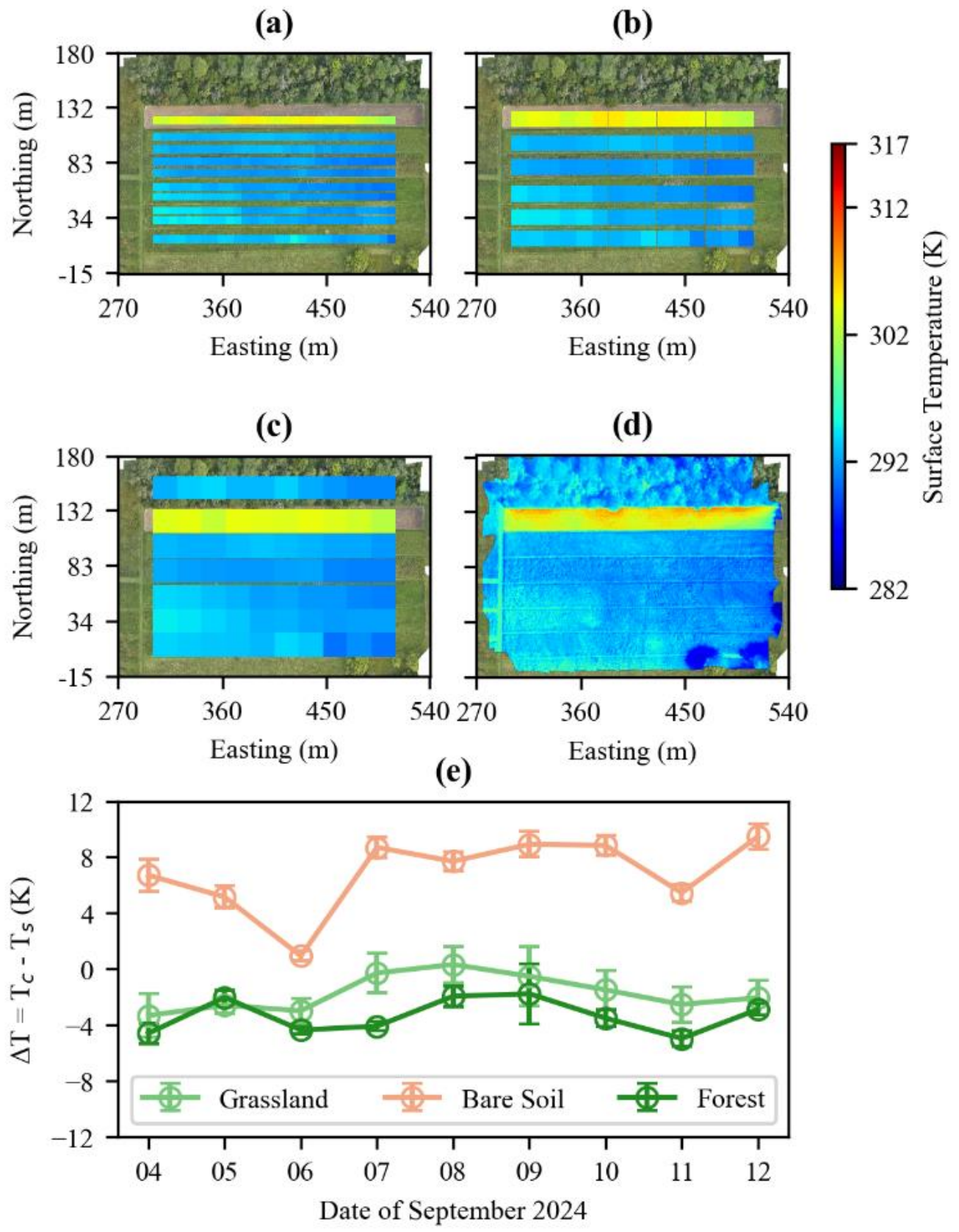

**Fig. 3**. Spatial distributions of canopy and soil surface temperatures measured by the DJI Mavic 3T at the grid cells of (a) 7 m, (b) 14 m, (c) 21 m, and (d) 0.1 m on September 8, 2024; (e) time series of temperature differences between canopy and soil surface temperatures ( $T_c$ ) and soil temperatures measured at a depth of 5 cm ($T_s$) across land covers.

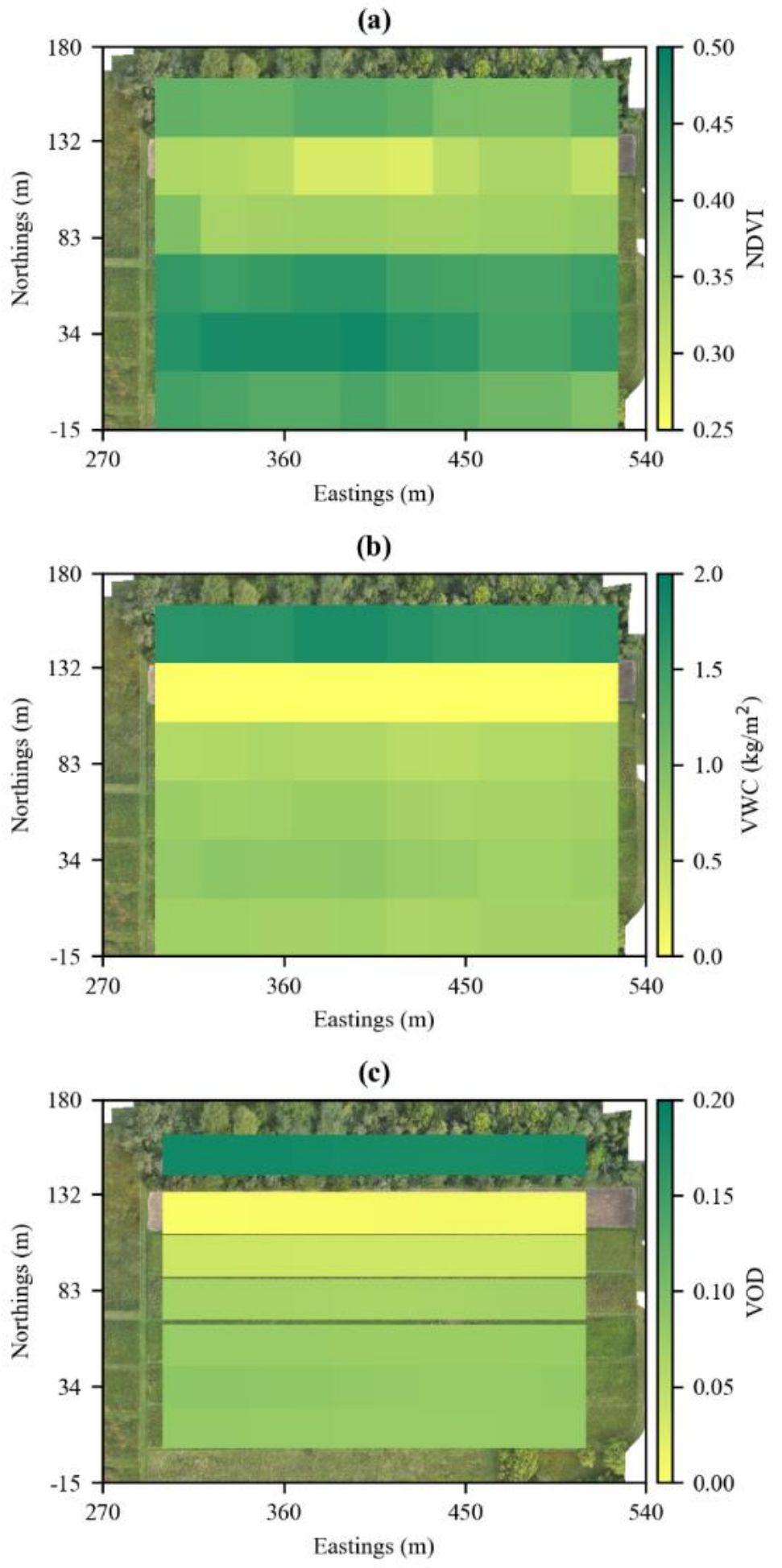

**Fig. 4.** Spatial maps of (a) Landsat 8 30 m Normalized Difference Vegetation Index (NDVI); (b) NDVI-derived vegetation water content (VWC); and (c) gridded vegetation optical depth (VOD) at a resolution of 21 m during the experimental period.

As mentioned above, accurate estimation of VOD (or τ) is imperative for SCA and is often derived from multi-spectral vegetation indices, including NDVI and the Leaf Aera Index (LAI) [51, 55]. In the SMAP retrieval framework, for example, VOD (or $\tau$ ) is computed as the product of a vegetation-related parameter ( $b_{LC}$ ) and vegetation water content (VWC), with VWC derived from a long-term NDVI climatology with land-cover-dependent coefficients (17) [9]. Although the use of NDVI climatology-based τ cannot fully capture real-time vegetation dynamics and can suppress short-term SM anomalies, the bulk precision of SMAP SCA SM retrievals remains comparable (or even better) to that of other algorithms [56, 57]. Given the short duration of this study's experimental period, the impact of using fixed NDVI-derived $\tau$ values on PoLRa SCA retrieval accuracy is expected to be limited. Surface reflectance data from red and near-infrared bands of Landsat 8 Level 2 and the Sentinel-2 Level 2 data sets (October 1, 2023 – September 30, 2024) have been used to compute maximum, minimum and effective NDVI values at 30 m and 10 m resolutions [58, 59]. These NDVI maps were first converted into VWC and VOD using (16), and the derived VOD matrices were subsequently resampled through spatial averaging to match the target retrieval resolutions (**Fig. 4** and **Fig. S1**). Preliminary comparisons indicated that the PoLRa SCA SM retrievals based on Landsat 8 and Sentinel-2 VOD products have similar accuracy to each other. SCA SM retrievals based on Landsat 8 VOD product have been solely adopted for subsequent analyses.

$$\mathrm{VWC} = (1.9134 \cdot \mathrm{NDVI}_{\mathrm{eff}}^2 - 0.3215 \cdot \mathrm{NDVI}_{\mathrm{eff}}) + f_{\mathrm{stem}} \cdot \frac{\mathrm{NDVI}_{\max} - \mathrm{NDVI}_{\min}}{1 - \mathrm{NDVI}_{\min}} \quad (16)$$

$$\tau = b_{\mathrm{LC}} \cdot \mathrm{VWC} \quad (17)$$

### *E. Bayesian Inference and Markov Chain Monte Carlo Simulation*

Bayesian inference is a statistical approach that uses Bayes's theorem to update prior beliefs based on observed data, yielding posterior probabilities for unknown parameters [40, 60]. In this study, we used Bayesian inference to probabilistically estimate SM and VOD from observed brightness temperatures, i.e., $T_{B_p}^{grid}$ . According to Baye's theorem, the posterior probability distribution, $\Pr(\theta|T_{B_p}^{grid})$, of the unknown parameters, $\theta = [\mathrm{sm}, \mathrm{VOD}]$, is computed by updating prior knowledge $\Pr(\theta)$ with the likelihood function $\Pr(T_{B_p}^{grid}|\theta)$ (18). In (18), the normalizing term $\Pr(T_{B_p}^{grid})$ is commonly ignored, and thus the posterior distributions of the unknown parameters can be inferred using (19). In this study, uniform priors within fixed pragmatically defined ranges of SM and VOD are used. The likelihood function is modeled in logarithmic form assuming that the error between the radiometric simulations (i.e., $T_{B_p}^{sim}$) and observations (i.e., $T_{B_p}^{grid}$) is Gaussian distributed with zero mean and a variance of $\sigma^2$ (20). Due to the complexity and nonlinearity of the problem, the Markov Chain Monte Carlo (MCMC) sampling technique is employed to obtain the posterior distribution of $\theta = [\mathrm{sm}, \mathrm{VOD}]$. Specifically, an ensemble MCMC sampler is implemented that iteratively explores the parametric spaces by generating multiple chains through the probabilistic transitions, converging towards the posterior distribution. The resulting MCMC chains provide posterior probabilistic predictions of SM and VOD. The Maximum-A-Posteriori (MAP) value or mode of a posterior distribution indicates the most probable value of the parameters, and the standard deviation of the posterior distribution has been used to represent retrieval uncertainty.

$$\Pr\left(\theta|T_{B_p}^{grid}\right) = \frac{\Pr\left(T_{B_P}^{grid}\middle|\theta\right)\Pr(\theta)}{\Pr\left(T_{B_P}^{grid}\right)} \quad (18)$$

$$\Pr\left(\theta|T_{B_P}^{grid}\right) \propto \Pr\left(T_{B_P}^{grid}\middle|\theta\right)\Pr(\theta) \quad (19)$$

$$\Pr\left(T_{B_P}^{grid}\middle|\theta\right) = -\frac{1}{2}\left(\frac{\left[T_{B_P}^{sim}(\theta) - T_{B_P}^{grid}\right]^2}{\sigma^2} + \log(2\pi\sigma^2)\right) \quad (20)$$

### *F. Validation Metrics*

Conventional evaluation metrics, including bias, unbiased root-mean-square error (ubRMSE), and Pearson correlation coefficient (R), have been utilized to assess the accuracy of PoLRa derived SM retrievals, as shown from (21) to (24) [61]. *in situ* SM measurements from TEROS 12 probes were used as the validation benchmarks after undergoing a correction scheme. Specifically, SM estimates from the TEROS 12 probes were compared with gravimetric SM samples collected on September 13, showing a strong linear correlation (**Fig. S2**). Based on this relationship, a linear correction function was derived to rescale the probe measurements to match the gravimetric SM magnitudes. This correction function was then applied to adjust probe measurements from other days, and the resulting rescaled SM values were used as benchmarks for evaluating PoLRa-derived SM retrievals. The ubRMSE quantifies discrepancies in magnitude between the retrieved and benchmarked SM, while R measures the varying consistency between the two datasets.

$$\mathrm{bias} = E[\mathrm{sm}_{\mathrm{ret}}] - E[\mathrm{sm}_{\mathrm{ref}}] \quad (21)$$

$$\mathrm{RMSE} = \sqrt{E[(\mathrm{sm}_{\mathrm{ret}} - \mathrm{sm}_{\mathrm{ref}})^2]} \quad (22)$$

$$\mathrm{ubRMSE} = \sqrt{\mathrm{RMSE}^2 - \mathrm{bias}^2} \quad (23)$$

$$R = \frac{E[(\mathrm{sm}_{\mathrm{ret}} - E[\mathrm{sm}_{\mathrm{ret}}])(\mathrm{sm}_{\mathrm{ref}} - E[\mathrm{sm}_{\mathrm{ref}}])]}{\sigma_{\mathrm{ret}}\sigma_{\mathrm{ref}}} \quad (24)$$

where E […] represents the arithmetic mean; the subscript obs and ref denote soil moisture retrievals from PoLRa observations and the spatially averaged *in situ* measurements; $\sigma_{\mathrm{ret}}$ and $\sigma_{\mathrm{ref}}$ refer to the standard deviations of the soil moisture retrievals and benchmarks.

## IV. RESULTS AND DISCUSSION

### *A. Brightness Temperature Characteristics Across Land Covers and Resolutions*

The collective distributions of PoLRa-observed brightness temperatures at individual footprints ($T_{B_p}$) have been analyzed and compared across land cover classifications and spatial resolutions. To ensure a fair comparison, low-resolution $T_{B_p}$ observations collected over the forested portion have been excluded from **Fig. 5a**. Using $T_{B_V}$ data as an example, an important feature is the increased spread in $T_{B_V}$ values with decreasing flight altitudes, likely reflecting the smoothing effects of coarser-resolution observations on within-pixel variability [18]. Overall, $T_{B_V}$ distributions at multiple resolutions exhibit similar magnitudes and peak patterns around 240 K, although at medium resolution, the $T_{B_V}$ peak shift slightly towards higher values compared to those at low and high altitudes (**Fig. 5a**). In **Fig. 5b**, $T_{B_V}$ observations at a resolution of 21 m are categorized by land cover class, showing a decreasing trend in $T_{B_V}$ values with increasing vegetation density. Given that the flight paths are predominately influenced by shrubland characteristics, leading to overlapping peaks between the $T_{B_V}$ distributions from all strips and those from shrubland only.

Beyond aggregated distributions, footprint-scale $T_{B_p}$ values from different resolutions have also been compared along flight tracks. Observations over Shrub 5 and Soil strips on September 8 were selected to illustrate resolution-dependent features. Given the rather dense PoLRa sampling, angular corrected $T_{B_p}$ data were first binned every one meter along the horizontal direction, and the mean $T_{B_p}$ values within each bin formed along-track values. Over Shrub 5, $T_{B_V}$ values across three resolutions are highly consistent, varying by less than 2 K along the flight line (**Fig. 5c**). In contrast, over bare soil, larger fluctuations in $T_{B_V}$ have been observed, which are consistently captured at all resolutions (**Fig. 5d**). Mid- and high-altitude $T_{B_V}$ values are generally closer than those at low altitude, likely reflecting the closer flight time windows at 20 m and 30 m. Nonetheless, the discrepancies among resolutions mostly remain within 5 K, corresponding to SM differences of approximately 2%. While the above results are solely based on $T_{B_V}$ data, $T_{B_H}$ observations have shown similar patterns, although the gradients in $T_{B_H}$ data across vegetation densities seem less pronounced (**Fig. S3**).

These multi-resolution footprint comparisons highlight expected changes across vegetation types. To further investigate spatial variability, gridded $T_{B_p}$ maps at resolutions of 7 m, 14 m, and 21 m on September 9 have been illustrated as an instance (**Fig. 6**). As expected, $T_{B_p}^{grid}$ fields become smoother with increasing flight altitude, yet key spatial features are preserved across resolutions and polarizations. $T_{B_p}^{grid}$ values are generally higher over Soil and lower over Shrub 5. Spatial patterns within shrubland strips show subtle gradients, and $T_{B_p}^{grid}$ values tend to be higher on the left side on September 9 (**Fig. 6**). Daily spatial correlations between $T_{B_V}^{grid}$ and $T_{B_H}^{grid}$ maps are 0.67, 0.74, and 0.52 at 7 m, 14 m, and 21 m resolutions, respectively. The reduced spatial correspondence at 21 m could be attributed to contrasting polarization behavior in the forest strip and associated calibration errors. Another notable distinction in polarized $T_B$ values is their dynamic ranges where $T_{B_H}^{grid}$ values exhibit approximately twice the spread of $T_{B_V}^{grid}$. Comparing $T_{B_p}^{grid}$ data at medium (14 m) and low (21 m) resolutions over Shrub 2 and Shrub 3 where 7 m pixels are fully nested within larger grids shows that coarser-resolution $T_{B_p}^{grid}$ values closely approximate the spatial averages of finer-resolution observations at 7 m, consistent with previous findings [18, 62].

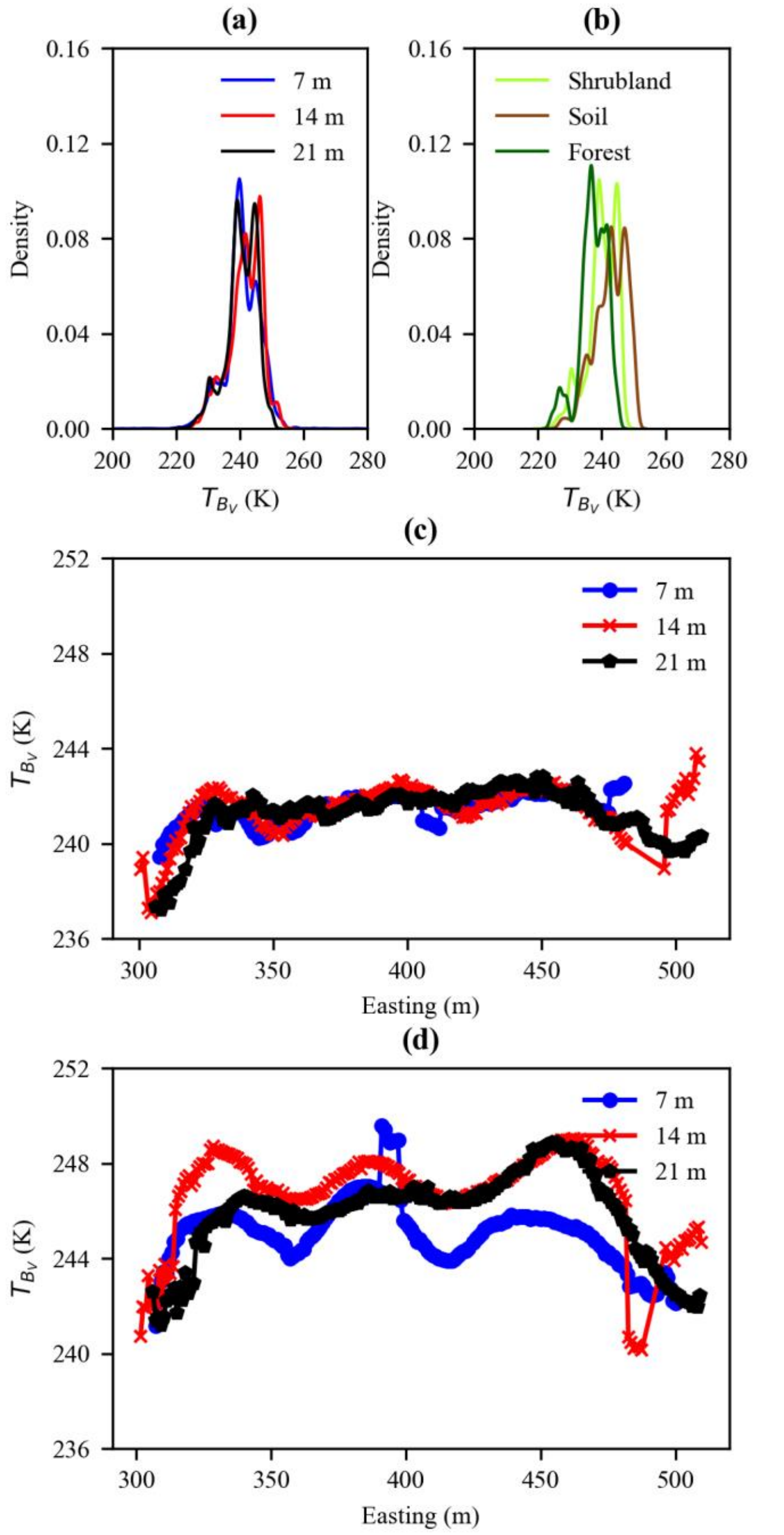

**Fig. 5.** (a) Probability density functions of all the vertically-polarized footprint-scale brightness temperatures collected over the experiment period at the grid scales of 7 m, 14 m, and 21 m; (b) probability density functions of 21 m vertically-polarized footprint-scale brightness temperatures conditioned by land covers; (c) variations of vertically-polarized footprint-scale brightness temperatures at resolutions of 7 m, 14 m, and 21 m over Shrub 5 on September 8, 2024; (d) variations of vertically-polarized footprint-scale brightness temperatures at resolutions of 7 m, 14 m, and 21 m over bare soil on September 8, 2024.

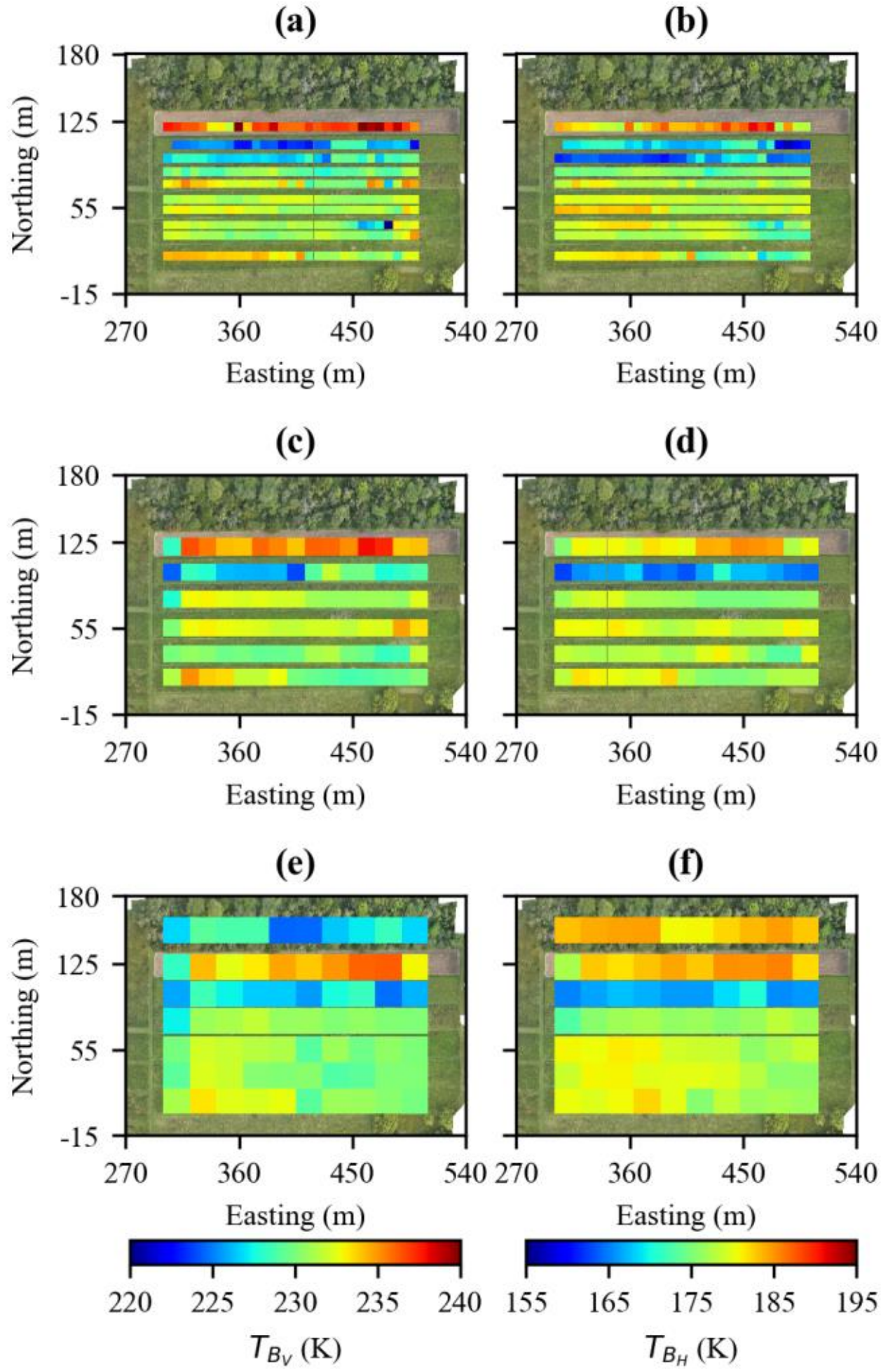

**Fig. 6.** Spatial maps of vertically polarized brightness temperatures at resolutions of (a) 7 m, (c) 14 m, and (e) 21 m; and spatial maps of horizontally polarized brightness temperatures at resolutions of (b) 7 m, (d) 14 m, and (f) 21 m on September 9, 2024.

*B. Performance Analysis of PoLRa-based Soil Moisture Estimates*

Multiple algorithms based on the identical $\tau - \omega$ framework have been applied to derive SM from PoLRa-observed $T_{B_p}^{grid}$ values at different altitudes. Although the magnitudes of SM retrievals vary across different algorithms, their spatial distributions generally follow the spatial patterns observed in $T_{B_p}^{grid}$ data. For example, both $T_{B_p}^{grid}$ and SM retrieval maps on September 9 display drier conditions over bare soil and a distinct east-west gradient, with drier soils in the east and wetter conditions in the west (**Fig. 7**). It should be noted that the spatial patterns of SM are temporally variant. On September 4, an opposite trend, a wet east and dry west, is observed (**Fig. S4**). Nevertheless, dry-to-wet SM gradients across soil, shrubland, and forest remain consistently evident across different days. Additionally, the relatively wetter tendency of Shrub 5 on September 9 has also been captured in multiple days. The reduced contrast between SM retrievals in Shrub 5 and other shrub strips, compared to the corresponded $T_{B_p}^{grid}$ patterns, may be attributed to the smoothing effects of the broad SM range and to the lower vegetation density in Shrub 5 compared to other strips. Under similar radiation intensities, denser vegetation would enhance attenuation effects and generally results in wetter SM retrievals, while sparse canopies lead to relatively drier estimates. This sensitivity is influenced by land cover type and surface conditions. Supporting this, significantly higher SCAH SM values have been obtained in the forest region than those in bare soil, despite their comparable $T_{B_H}^{grid}$ values on September 9. Systematic dry DCA SM estimates on September 9 are associated with unreasonably low VOD retrievals, which could be caused by the compensation mechanism under the DCA algorithm and the original errors in $T_{B_p}^{grid}$ measurements and/or calibration. In contrast, DCA retrievals on September 4 produce higher SM estimates than SCA's outputs, corresponding to the elevated VOD magnitudes that mostly exceed those derived from NDVI values. When using the same VOD inputs, differences in polarized $T_{B_p}^{grid}$ measurements and Fresnel reflectivity primarily account for the differences between SCAV and SCAH SM retrievals. The MT-DCA SM estimates generally fall between those retrieved by SCA and DCA approaches (**Fig. 7**).

The retrieval accuracy of PoLRa-derived SM estimates is evaluated by comparing retrievals to re-scaled probe measurements using statistical metrics, including ubRMSE and R (Table 2). In contrast to RMSE, ubRMSE that is less sensitive to biases in both mean and fluctuation amplitudes, has been commonly used to assess the relationship between retrievals and reference SM [10]. Independent of magnitude discrepancies, the correlation coefficient, R, measures the consistency of temporal evolutions between retrievals and reference measurements. In general, PoLRa retrievals tend to overestimate SM by approximately 0.1 m3/m3. Among the retrieval algorithms, ubRMSE values for SCAV and MT-DCA retrievals are generally below 0.04 m3/m3, meeting the SMAP mission's accuracy requirement [9]. Thus, subsequent analyses primarily focus on these two algorithms. In contrast, R values are typically lower than 0.45, largely due to the limited sample size across the 9-day experimental period, where minor anomalies could significantly affect temporal correspondence. The negative R values observed for SCAH retrievals likely indicate calibration errors or noise in $T_{B_H}^{grid}$ inputs. Nevertheless, the improved R values for DCA and MT-

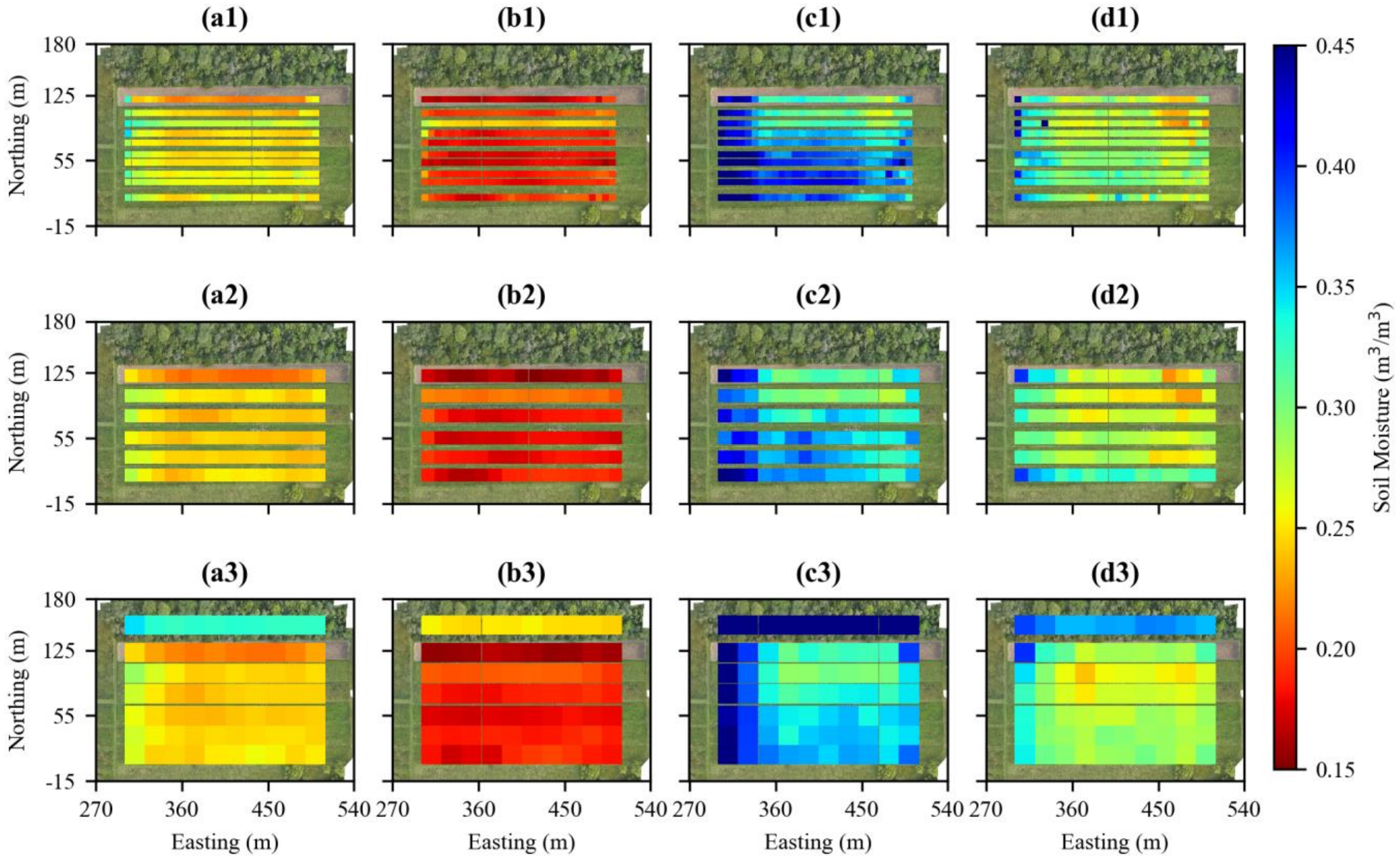

**Fig. 7.** Spatial distributions of soil moisture retrievals derived from (a1) – (a3): the single channel algorithm based on vertically polarized brightness temperatures; (b1) – (b3): the single channel algorithm based on horizontally polarized brightness temperatures; (c1) – (c3): the dual channel algorithm; and (d1) – (d3) the multi-temporal dual channel algorithm on September 9, 2024. The top, medium, and bottom rows represent the spatial resolutions of soil moisture at 7 m, 14 m, and 21 m, respectively.

DCA retrievals suggest that the use of dual-polarization observations from the same or adjacent days can partially mitigate the effects of noisy inputs and invariant VOD across the experimental period for SCA here (**Fig. S5**). Overall, MT-DCA retrievals exhibit the best integrated performance in terms of ubRMSE and R under the conditions of this field campaign.

Table 2. Evaluation metrics of PoLRa-derived SM retrievals using different algorithms

| Algorithm | Resolution (m) | Bias ($m^3/m^3$) | RMSE ($m^3/m^3$) | ubRMSE ($m^3/m^3$) | R |
|---|---|---|---|---|---|
| SCAV | 7 | 0.088 | 0.094 | 0.031 | 0.257 |
| | 14 | 0.077 | 0.083 | 0.029 | 0.324 |
| | 21 | 0.103 | 0.110 | 0.036 | 0.207 |
| SCAH | 7 | 0.096 | 0.107 | 0.045 | -0.146 |
| | 14 | 0.077 | 0.091 | 0.047 | -0.144 |
| | 21 | 0.107 | 0.123 | 0.059 | -0.181 |
| DCA | 7 | 0.100 | 0.110 | 0.044 | 0.451 |
| | 14 | 0.094 | 0.106 | 0.045 | 0.464 |
| | 21 | 0.114 | 0.127 | 0.054 | 0.404 |
| MT-DCA | 7 | 0.092 | 0.099 | 0.035 | 0.403 |
| | 14 | 0.089 | 0.098 | 0.037 | 0.441 |
| | 21 | 0.108 | 0.116 | 0.041 | 0.422 |

Among different algorithms, ubRMSE values markedly increase at the 21 m resolution compared to 7 m and 14 m, with the performance degradation primarily driven by poorer retrievals over the forest regions (**Fig. 8**). Considering VWC distributions from Landsat 8 NDVI data and using a threshold of 0.04 $m^3/m^3$, PoLRa retrievals appear more accurate when VWC is below 1.5 $kg/m^2$. This preference could also explain the reduced accuracy of SCA retrievals based on Sentinel-2 VOD products where mean VWC values are 1.2 $kg/m^2$ in shrubland and 2.5 $kg/m^2$ in forest (**Fig. S1**). In areas with VWC below this threshold, accurate representation of VOD is critical for SM retrieval precision. For MT-DCA retrievals, ubRMSE values over shrublands are approximately 0.03 $m^3/m^3$, slightly better than those over bare soils (**Fig. 8**). In contrast, for SCAV retrievals, ubRMSE values were consistently smallest over bare soil (~ 0.02 $m^3/m^3$) across all land cover types, likely benefiting from the assumption of zero VOD over bare soil. Both SCAV and MT-DCA retrievals demonstrate comparable accuracy over shrubland strips, although MT-DCA outperforms in Shrub 3.

Example of SM time series from PoLRa retrievals and *in situ* probe measurements for three representative pixels separately in Shrub 5, Soil, and Forest are presented in **Fig. 9**. Consistent with spatial patterns, SM magnitudes increase with vegetation density. Due to the absence of precipitation during the experiment, *in situ* probe measurements exhibit a general drying trend, although minor fluctuations have also been observed. In contrast, PoLRa retrievals show two distinct drying segments: September 5 – 8 and September 9 – 12.

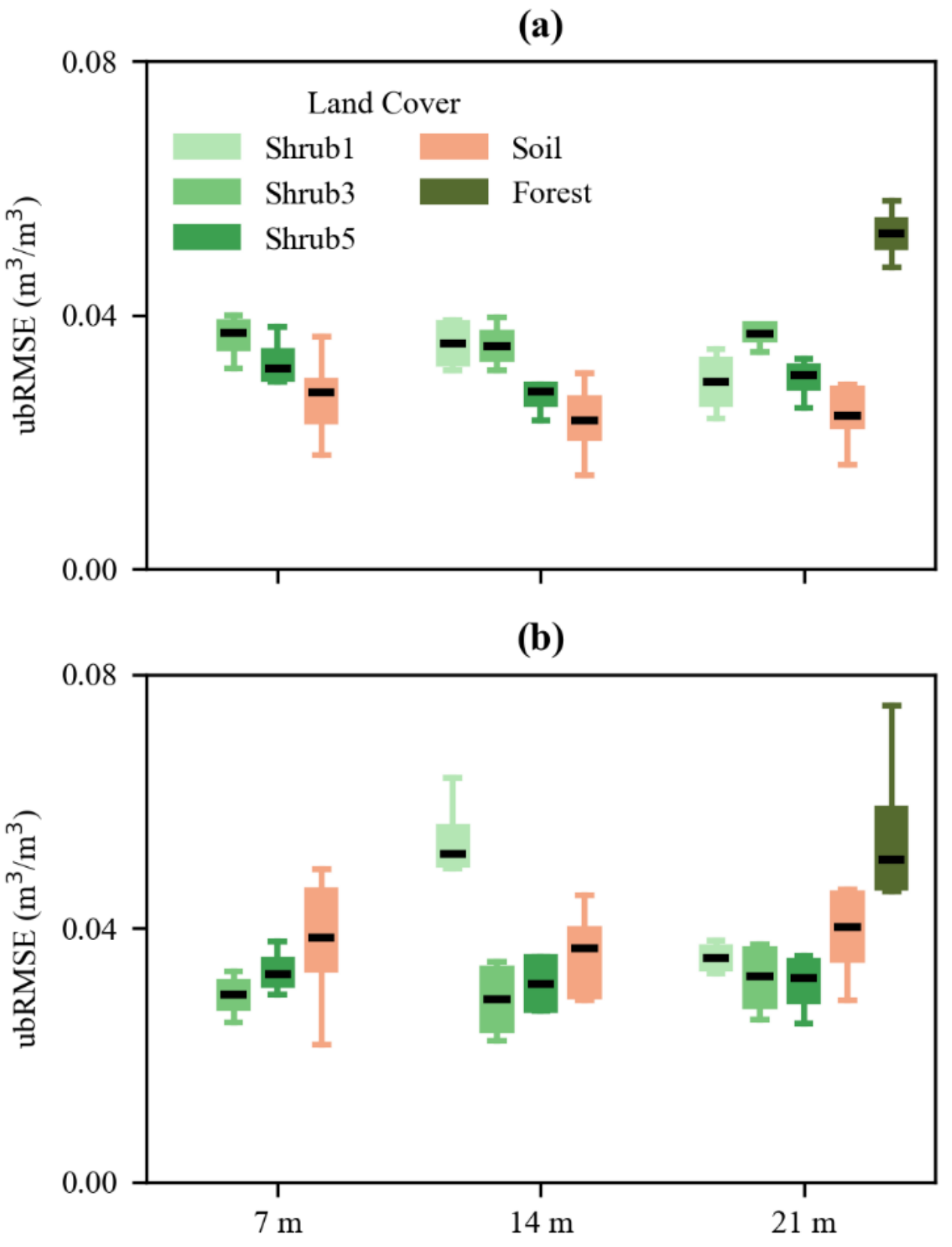

**Fig. 8.** Boxplots of ubRMSE values for soil moisture retrievals derived using (a) the single channel algorithm based on vertically polarized observations and (b) the multi-temporal dual channel algorithm across different land covers.

While unexpected rises between these periods may reflect calibration artifacts, similar minor increases are also observed in bare soil *in situ* data (**Fig. 9**). Consistent with prior studies, L-band radiometer-based SM time series has larger amplitude variations than those measured by *in situ* probes [63].

### C. Bayesian Uncertainty of PoLRa Retrievals

As mentioned earlier, Bayesian inference produces full posterior distributions of SM and VOD rather than deterministic point estimates obtained by the least-squares optimization. The MAP estimate, the mode of the posterior distribution, represents the most probable value of SM and/or VOD given the observed $T_{B_p}^{grid}$ values and prior information. Ideally, in this study, the MAP estimates and the least-squares solutions should converge as both aim to minimize the differences between modelled and observed $T_B$ values. Such a convergence can be observed for the MAP and least-squares SM retrievals from the SCA algorithm (**Fig. 10a-b**). For the DCA-based retrieval frameworks, however, an average RMSE of approximately 0.01 $m^3/m^3$ indicates the existence of small discrepancies (**Fig. 10c** and **Fig. 10e**). These differences in the context of DCA-related algorithms are expected to alleviate when the bounds on SM and VOD are extended, such as **Fig.S6**. Yet, retrievals that exceed physically plausible limits could suggest (1) significant inconsistencies between dual-polarized

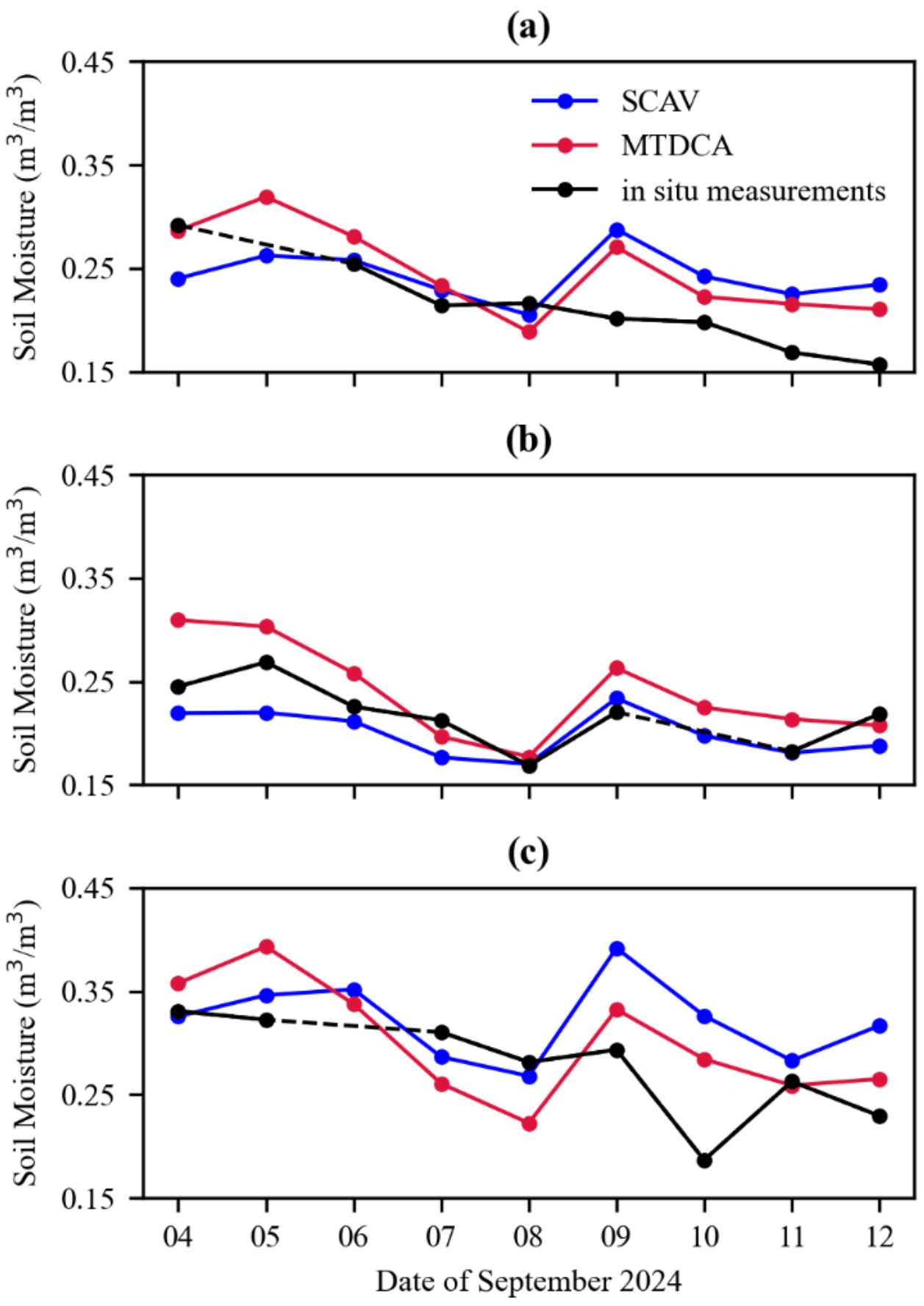

**Fig. 9.** Time series of *in situ* soil moisture measurements (black), and PoLRa-derived retrievals using the single channel at vertical polarization (blue) and the multi-temporal dual channel algorithm (red) in (a) Shrub 5, (b) Soil, and (c) Forest.

observations and/or (2) unrealistic $T_{B_p}^{grid}$ measurements. Given the overall quality of SCA SM retrievals, the second explanation, unrealistic $T_{B_p}^{grid}$ measurements, is less likely, and mismatched dual-polarized $T_{B_p}^{grid}$ values induced by calibration and/or instrument issues are more plausible contributors. When optimal solutions fall outside prescribed physically reasonable bounds, least-square retrievals often drive the VOD estimate to zero, whereas MAP estimates are less prone to cluster at zero (**Fig. 10d** and **Fig. 10f**). As a result, MAP SM estimates tend to be slightly higher than their deterministic counterparts (**Fig. 10c** and **Fig. 10e**). The retrievals likely affected by this mechanism have been highlighted as orange points shown in **Fig. 10c-f**. Nonetheless, these differences are rather trivial and do not necessarily degrade SM retrieval accuracy, as observational noise tends to impact VOD retrievals more severely than SM estimates (e.g., **Fig. 10f**) [64]. In practice, abrupt drops of VOD to zero over time are not uncommon for DCA-based retrievals, including the SMAP official SM product where an additional regularization term is also included in the retrieval objective function [33]. Furthermore, the full posterior distributions generated by

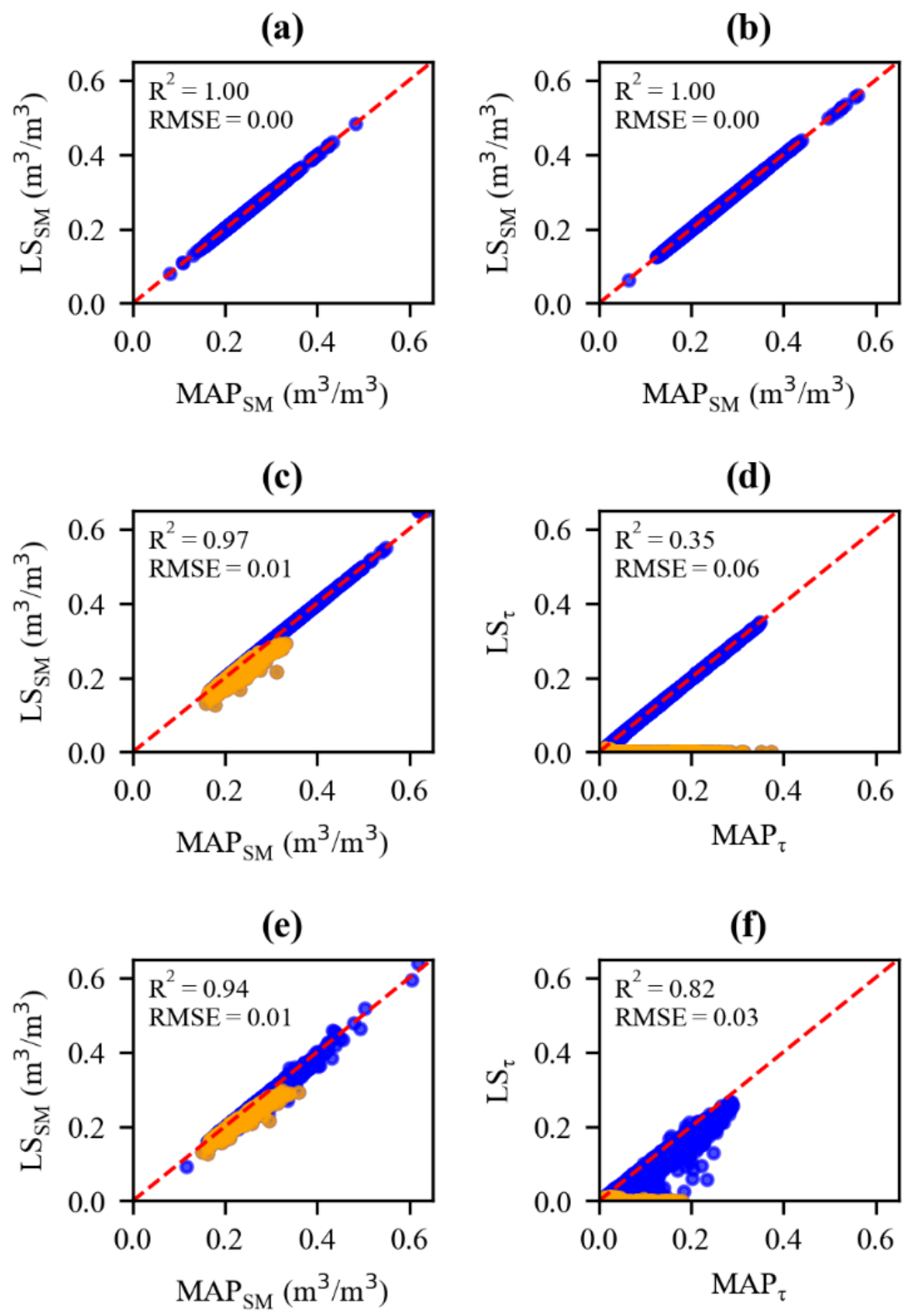

**Fig. 10.** Scatterplots of soil moisture (SM) and vegetation optical depth ($\tau$) retrievals derived from the least-square optimization (LS) and the maximum-a-posterior (MAP) based on Bayesian inference in the context of the single channel algorithm (a) at the vertical polarization and (b) horizontal polarization, the dual channel algorithm (c-d), and multi-temporal dual channel algorithm (e-f).

Bayesian inference can sometimes capture cases where non-unique DCA solutions may exist. Despite these, MAP and least-square estimates generally converge closely.

The standard deviations derived from the posterior distributions have been used in this study to represent the uncertainty spreads of PoLRa-retrieved SM and VOD values. Although the posterior mean and the MAP estimates are mostly not aligned, the standard deviation surrounding the MAP estimate is typically considered equivalent to that around the mean [65]. It is important to note that, in the absence of parameter uncertainty, retrievals from the SCA algorithm produce a unique deterministic solution, and the corresponding uncertainty intervals generated through Bayesian inference are extremely narrow, on an average magnitude of 0.02 m$^3$/m$^3$. This is because the error variance ($\sigma^2$) in (20), associated with a single $T_B$, is set to $10^{-6}$, which causes the sampler to explore a very narrow range and results in sharply peaked posterior distributions. Meanwhile, the mean standard deviations for DCA- and MT-DCA-retrieved SM are approximately ± 0.12 m$^3$/m$^3$ and ± 0.11 m$^3$/m$^3$, respectively. Rather than being fixed, $\sigma^2$ values in (20) for DCA and MT-DCA algorithms are dynamically scaled and impact the following sampling pools. Given this, direct comparisons of standard deviations or uncertainty ranges across various algorithms are unfair. When no perfect SM-VOD combinations that yield zero residuals for balancing polarized $T_{B_p}^{grid}$ observations, exist in the pre-defined search spaces, a “cloud” of suboptimal solutions emerge, which significantly inflates the uncertainty range. These large residuals could be attributed to simulation-observation mismatches induced by imperfect model and improper parameters (in addition to SM and VOD), misaligned observations across polarizations due to sensor anomaly and calibration deviations, and other reasons, which are beyond the scope of this work. **Fig. 11.** The spreads of 21 m soil moisture uncertainty are based on (a) the single channel algorithm, (b) the dual channel algorithm, and (c) the multi-temporal dual channel algorithm.**Fig. 11** illustrates the spreads of 21 m SM uncertainty ranges based on 9-day MAP averages of different algorithms at 28 locations where probe measurements were conducted. The reference SM values largely fall within the one-fold uncertainty intervals (i.e., ± 1$\sigma$), especially over bare soil, i.e., site 12 - 20 (**Fig. 11**).

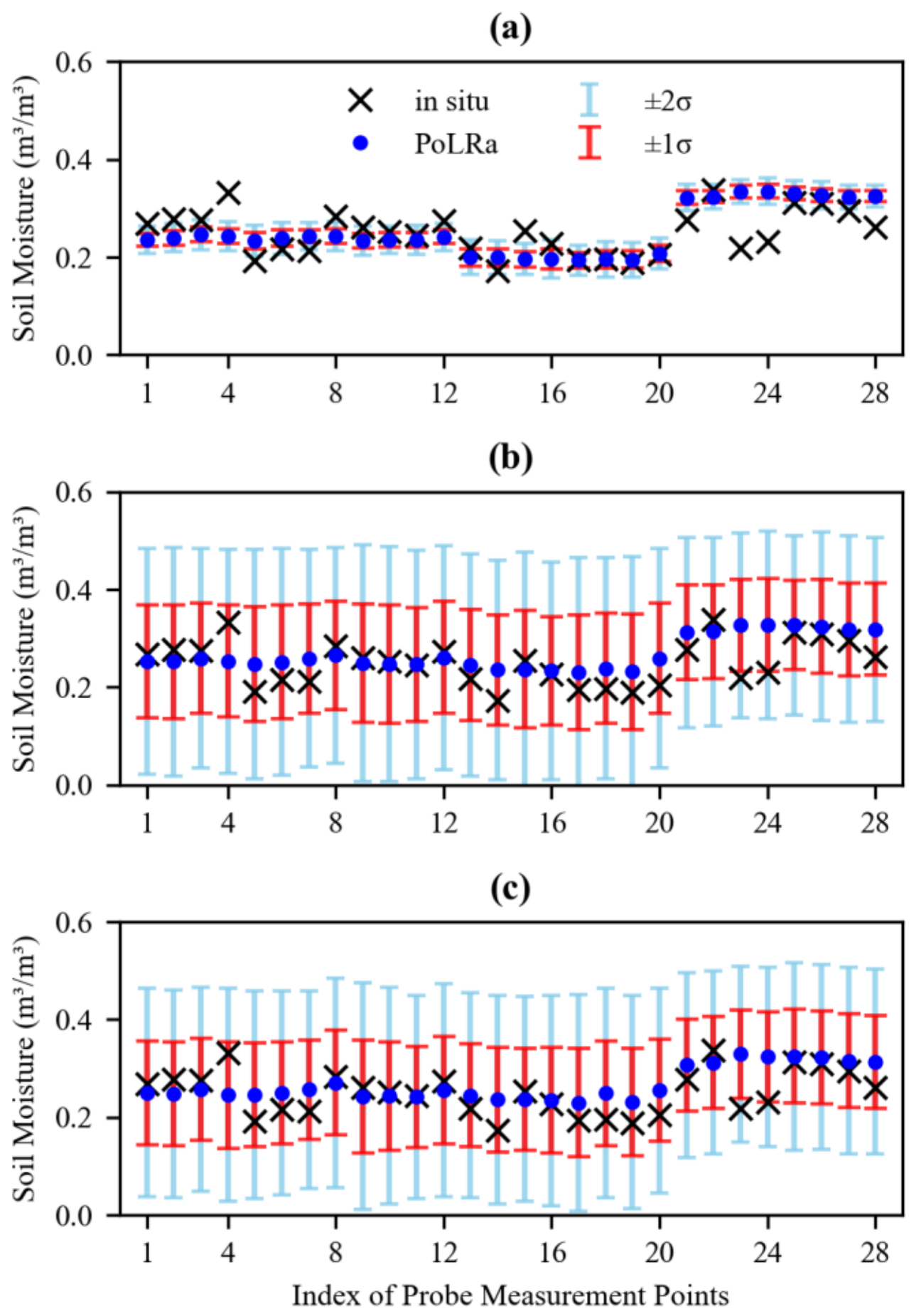

**Fig. 11.** The spreads of 21 m soil moisture uncertainty are based on (a) the single channel algorithm, (b) the dual channel algorithm, and (c) the multi-temporal dual channel algorithm.

### *D. Discussion*

Due to the lack of a standardized calibration manual and emission benchmark, one of the primary challenges in broadly deploying drone-based radiometers for SM and vegetation sensing lies in accurately characterizing the ACS to obtain reliable $T_{B_p}$ data. Although cold-sky measurements were conducted daily during the experimental period, the manual setup could not guarantee consistent viewing geometries across days, and the magnitudes of daily offsets used to counter impedance mismatch and/or low-magnitude background interference are not negligible. In terms of SM accuracy, applying uniform daily offsets does not improve agreement between PoLRa retrievals and *in situ* probe measurements. However, the consistent spatial patterns of dual polarization $T_{B_p}$ observations across spatial resolutions partially support the validity of the daily calibration parameters. The reduced R values, particularly for horizontally polarized retrievals, may imply the calibration parameters may have varied inappropriately over time. Indeed, opposite trends in calibrated $T_{B_H}$ and $T_{B_V}$ measurements across successive days were observed. Incorporating partial *in situ* measurements into the calibration process to produce daily offsets similar in magnitude to those obtained from sky measurements could improve R values. However, these ground-based offsets have inadvertently captured SM trends, thereby biasing evaluation metrics. Therefore, consistent radiation reference measurements, such as those from sky or inland water bodies, are critical for reliable calibration. To obtain two "free-space" calibration sources at the antenna input, as opposed to at the interval switch, the inclusion of a microwave absorber with known $T_{B_p}$ values for "warm" calibration is also recommended and could improve calibration accuracy [18].

On the algorithmic side, the unavailability of fine-scale model parameters (e.g., h and ω) complicates the retrieval of high-precision SM estimates. Land cover-dependent parameters have been adopted for satellite missions such as SMAP [9]. Although these empirical values have initially derived from field campaigns, recent studies have shown that ω can vary more within a given land cover class than between different classes [38]. Furthermore, h and ω parameters are known to vary with landscape conditions, temporal dynamics, and algorithm type [66, 67]. While SM retrievals based on h of 0.13 and ω of 0.08 have exhibited small ubRMSE values, these choices are likely suboptimal and cannot generalize across all land covers. Another critical parameter for SCA-based retrievals is the $b_{LC}$ coefficient, which translates VWC into VOD, and plays a key role during active vegetation growth periods [17]. Although temporal fluctuations of these parameters are likely insignificant in this experiment with a 10-day duration, they must be carefully considered for studies over longer periods. Additionally, the spatial mismatch between *in situ* SM benchmarks and gridded retrievals could have affected the validation results. To maximize the evaluation scenarios, a number of *in situ* measurements outside the pixels were still used as benchmarks (for their most-adjacent grids). Minor and negative correlations between SCAH retrievals and probe measurements suggest that a fraction of $T_{B_H}$ inputs are likely biased, possibly due to errors in measurements and/or daily calibration. Despite the potentially unexpected biases in $T_{B_H}$ values, SM retrievals from the MT-DCA algorithm still achieved the lowest ubRMSE values compared to algorithms relying on $T_{B_H}$ observations. This could indicate that MT-DCA retrievals are more tolerant to observational noise and model imperfections, consistent with previous findings [38].

Consistent with earlier results, drone-based PoLRa measurements have demonstrated good capability in retrieving surface SM over relatively homogeneous field [29, 30]. Future work will focus on characterizing SM variations across areas with heterogeneous land covers which become increasingly prevalent with higher flight altitudes. Although earlier studies suggest that subpixel heterogeneity minimally affects L-band observations, these conclusions are largely based on simulations and areas with gradual vegetation transitions [18, 62]. UAV-mounted radiometers, with their flexibility and high spatial resolution, provide a promising platform to systematically investigate heterogeneity effects at relatively low cost, when can help refine retrieval algorithms for satellite missions. Regarding SM retrieval uncertainty, the feasibility of applying Bayesian inference to drone-based PoLRa observations has been examined using $\tau - \omega$-based algorithms. Bayesian inference provides probabilistic retrievals instead of single-value deterministic estimates — enabling determination of the confidence or uncertainty of the retrievals. It also provides additional insights by depicting the near-optimal solutions in the posterior distributions of the retrieved parameters. Following this, the current framework could be extended by incorporating prior distributions for model parameters with both SCA and DCA schemes. However, it should be noted that fully simultaneous retrievals of optimal parameter values and their associated uncertainties would require richer observational information. For drone-based systems, this could potentially be achieved by adjusting flight configurations to collect multi-angular observations over short windows. It is also important to note that MAP estimates are effective parameters optimized for retrieval purposes and may not always reflect their true physical meanings [68].

## V. Conclusion

This study conducted a comprehensive evaluation of soil moisture (SM) retrievals using the Portable L-band radiometer (PoLRa) mounted on an unmanned aerial vehicle (UAV) platform across diverse land covers and spatial resolutions. A series of controlled flights were performed over shrubland, bare soil, and forest strips, with data collected at multiple altitudes to generate gridded brightness temperatures ($T_{B_p}^{grid}$) at spatial scales of 7 m, 14 m, and 21 m. Footprint-scale and gridded (polarized) brightness temperatures ( $T_{B_p}$ ) were analyzed to assess resolution-dependent variability and to investigate the impacts of vegetation density on microwave emissions. The results reveal that, although coarser resolutions tend to smooth spatial variations, key land surface features were preserved across resolutions. Notably, $T_{B_p}^{grid}$ distributions show consistent dry-wet gradients associated with changes in vegetation density, and spatial correlations between vertical and horizontal polarizations were generally maintained across scales.

Building on the $T_{B_p}^{grid}$ observations, SM retrievals were carried out using multiple $\tau - \omega$-based algorithms, including the single channel algorithm (SCA), dual channel algorithm (DCA) and its derivative multi-temporal dual channel algorithm (MT-DCA). SM retrievals from the SCAV and MT-DCA algorithms achieve unbiased root-mean-square errors (ubRMSE) generally below 0.04 $m^3/m^3$, meeting the SMAP mission's target accuracy requirement. Among all retrieval methods, MT-DCA consistently produces the best overall performance, with ubRMSE values around 0.03 $m^3/m^3$ and correlation coefficient (R) values around 0.45 over shrubland and bare soils. On the other hand, SCAH retrievals suffer from low or even negative R values, possibly highlighting calibration issues particularly at horizontal polarization. Compared to the DCA and SCAH, MT-DCA retrievals exhibit substantially improved robustness, demonstrating greater tolerance to measurement noise, calibration uncertainties, and imperfect model assumptions through the exploitation of multi-temporal observation constraints. SM retrieval accuracy of PoLRa is generally higher over areas with vegetation water content (VWC) less than 1.5 $kg/m^2$, and retrieval performance degraded over forested regions with denser vegetation. Bayesian inference results confirmed that reference SM measurements largely fall within the one-fold derived uncertainty intervals surrounding the maximum a posteriori (MAP) estimates. While the standard deviations surrounding the SCAV and SCAH MAP estimates approximate ± 0.02 $m^3/m^3$, the uncertainty ranges for DCA and MT-DCA are ± 0.12 $m^3/m^3$ and ± 0.11 $m^3/m^3$, potentially due to model-observation discrepancies and/or misaligned observations across polarizations. These findings highlight the strong potential of UAV-based PoLRa for high and scalable resolution SM sensing across multiple land cover conditions with continuous refinement on calibration techniques and standardization.

## Acknowledgment

The authors gratefully acknowledge James L. Ellis and Nathan C. Hudson for their support in providing and managing the experimental site. We also thank Daniel Jimenez Gil and Shamontee Aziz for their assistance with field data collection and soil analysis.

## AUTHOR INFORMATION

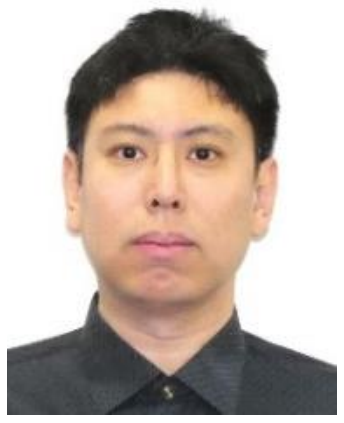

**Runze Zhang** received the B.S. degree in chemistry and the B.A degree in law (dual degree) from China Agricultural University, Beijing, China, in 2014, the M.S. degree in environmental engineering from University of New South Wales, Sydney, Australia, in 2018, and the Ph.D. degree in civil engineering from University of Virginia, Charlottesville, US, in 2023.

He is currently a Postdoctoral Researcher within the Department of Civil and Environmental Engineering at the University of Illinois at Urbana-Champaign. His research interests encompass the application of microwave remote sensing data on land hydrology, land surface emission modeling and the integration of unmanned aerial vehicle sensor data to investigate surface responses to wildfires.

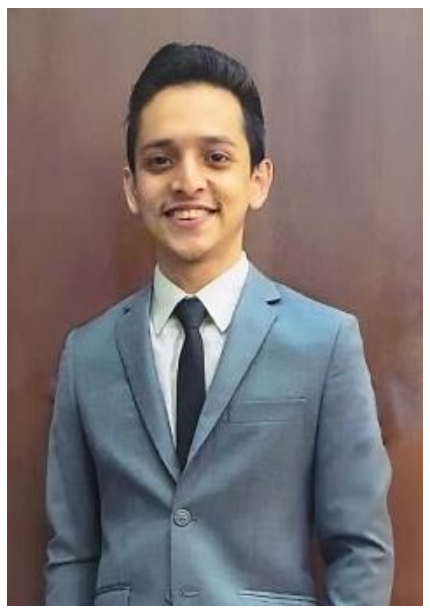

**Ishfaq Aziz** received his B.Sc. and M.Sc. degrees in Civil Engineering from Bangladesh University of Engineering and Technology (BUET), Dhaka, Bangladesh, in 2017 and 2020, respectively. Currently, he is pursuing a Ph.D. in Civil and Environmental Engineering at the University of Illinois Urbana-Champaign (UIUC), Urbana, IL, USA.

His research interest lies in radar sensing and nondestructive evaluation (NDE) of subsurface materials, forward modeling of radar wave propagation, and inverse parameter estimation using Bayesian Inference and Physics-Informed Machine Learning (PIML).

**Derek A Houtz** (Member, IEEE) was born in Washington DC, USA, Sept. 5, 1989, He has his B.S. and Ph.D. in aerospace engineering from the University of Colorado, Boulder, CO, USA (2011, 2017).

He currently works as a Research Scientist in the Microwave Remote Sensing group of the Swiss Federal Institute for Forest, Snow, and Lanscape Research (WSL), Birmensdorf, Switzerland. He is also the Co-Founder and C.E.O. of TerraRad Tech AG, Zurich, Switzerland. He previously worked as a post-doctoral researcher at the National Institute of Standards and Technology, Boulder, CO, USA. Dr. Houtz's research interests include radiometer calibration, radio-frequency electronics, remote sensing of the cryosphere, turfgrass soil moisture, precision irrigation, terrestrial radiometry and radar applications.

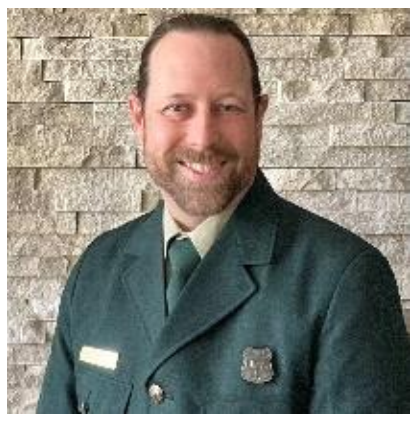

**Adam C. Watts** received the B.S. degree in biology and human and natural ecology (co-major) from Emory University, Atlanta, US, in 1999, and the M.S. and Ph.D. degrees in interdisciplinary ecology in 2002 and 2012, respectively, from University of Florida, Gainesville, US.

He was a Postdoctoral Research Ecologist in the School of Forest, Fisheries, and Geomatics Sciences at University of Florida from 2012 to 2013. He was Assistant and Associate Research Professor of Fire and Unmanned Systems at the Desert Research Institute (DRI), Reno, US, from 2013 to 2021, and Deputy Director of the Division of Atmospheric Sciences from 2019 to 2021 at DRI, Reno, US.

Dr. Watts is currently a Supervisory Research Biologist at the US Forest Service Pacific Wildland Fire Sciences Lab, Seattle, US. His research interests include systems and disturbance ecology, technology development and application, and fire science.

He is a member of the Association for Fire Ecology and the International Association for Wildland Fire.

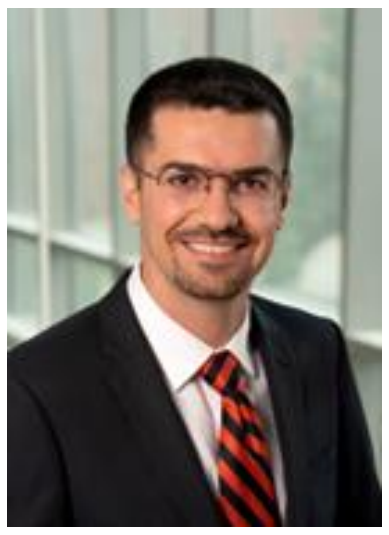

**Mohamad Alipour** received the the B.Sc. and M.Sc. degrees in civil engineering from Tehran Polytechnic, Tehran, Iran, in 2008 and 2011, respectively, and the Ph.D. degree in civil engineering from the University of Virginia, Charlottesville, VA, USA, in 2019.

He was a Postdoctoral Associate in the Department of Civil and Environmental Engineering at UCLA, Los Angeles, CA, USA, from 2020 to 2022. He is currently a Research Assistant Professor with the Department of Civil and Environmental Engineering, University of Illinois Urbana-Champaign (UIUC), and a Faculty Fellow of the National Center for Supercomputing Applications (NCSA), Urbana, IL, USA. His research interests include remote sensing and nondestructive evaluation, digital twins, and robotic infrastructure condition assessment.

Dr. Alipour is a member of the American Society of Civil Engineers (ASCE) and is currently an Associate Editor for the Journal of Nondestructive Evaluation.